\documentclass[amsmath,amssymb,aps,]{revtex4-2}

\usepackage{graphicx}
\usepackage{dcolumn}
\usepackage{bm}
\usepackage{verbatim}
\usepackage{xcolor}
\usepackage{ulem}
\usepackage{epstopdf}
\allowdisplaybreaks

\begin{document}
    \title{Global multimode squeezing in a train of ultrashort pulses from unbalanced SU(1,1) interferometers}
    \author{Wen Zhao$^1$}
    \author{Xiao Liu$^1$}
    \author{Yunxiao Zhang$^1$}
    \author{Xueshi Guo$^1$}
    \email{xueshiguo@tju.edu.cn}
    \author{Xiaoying Li$^1$}
    \email{xiaoyingli@tju.edu.cn}
\affiliation{%
$^{1}$The State Key Laboratory of Precision Measurement Technology and Instruments,\\
College of Precision Instrument and Opto-Electronics Engineering,\\ 
Tianjin University, Tianjin 300072, P. R. China
}
\date{\today}
             	
\begin{abstract} 
Time-domain multiplexed continuous-variable quantum states provide a promising route toward large-scale quantum networks. Existing platforms are based on continuous-wave pumped optical parametric systems, where the durations of temporal modes are on the order of nanoseconds. Here we demonstrate the time-domain multiplexed squeezing localized in a train of ultrashort pulses by exploiting unbalanced SU(1,1) interferometer (USUI) with a mode-locked laser serving as pump. Using the pulse-resolved measurement, we reveal the correlation structure of the state is unique and fundamentally different from previous approaches. To reach the ideal intensity squeezing, in principle, both the gain of USUI and mode number $M$ involved in joint measurement should tend to infinity, illustrating the feature of global multimode squeezing. We conduct proof-of-principle experiments, in which the temporal mode duration is down to 10 ps. 
We verify the intensity squeezing degree $R_d$ depends on both the gain of USUI and $M$. The results show $R_d$ improves with the increase of $M$ for $M<10$ and $R_d$ is lower than shot noise level by $\sim0.9$ dB for $M>10$ when the gain of USUI is fixed. 
Our investigations demonstrate the emission from high gain USUI is novel, which not only possesses the unique coherent feature but also enables the realization of ultra-large-scale quantum states.
\end{abstract}
\maketitle
	
\section{Introduction}
Quantum correlations are central to quantum science. Increasing the number of modes in quantum states can significantly enhance the capacity of quantum information systems. In the continuous variable (CV) quantum optics \cite{Weedbrook2012RevModPhys, Adesso2007JPA}, a variety of approaches based on high gain optical parametric amplifiers (OPAs) have been used to generate multi-mode quantum states correlated in different degrees of freedom including frequency, spatial, and temporal modes, etc. \cite{RevModPhys2020Treps, Pfister2013PRL, JiaXinyu2025Nat, Cai2017NatCommun, JingJietai2020PRL, Armstrong2012NatCommun, AndrewForbes2020SciAdv}. 
Among these schemes, time-domain multiplexing has proven to be an effective method for large-scale quantum state generation.
By coherently combining delayed CV Einstein-Podolsky-Rosen pairs generated from OPAs with linear beam-splitters or OPAs, large-scale CV entangled states have been generated, and their applications in quantum computation have been demonstrated \cite{menicucci2011temporal,Menicucci2011PRA,Yokoyama2013, Furusawa2019Science, Larsen2019Science, ZhouYanfen2023PRL,asavanant2021time}.  
	
Despite these developments, 
the advancement of quantum networks involving multi-user \cite{Bell2014, JiaXiaojun2024PhysRevRes, LiuShengshuai2020NatCommun} requires routing different modes into spatially separated location without reducing their quantum correlations. Therefore, precise time synchronization is a pivotal requirement for the time-domain multiplexed quantum state.
Existing work employs OPAs with single-frequency continuous wave laser serving as the pump, in which the correlated photon pairs are created randomly within the coherence time of pump. In this case, the durations of temporal modes are on the nanosecond scale. The synchronization for implementing the task of quantum information processing is defined by electronic gate with timing jitter on the order of tens picosecond. 
Therefore, the scale of the multimode quantum state is limited, and precise time synchronization required for quantum networks becomes challenging.
	
When the OPA is pumped by a mode-locked laser, correlated photons from the OPA are created within the pulse duration of pump, and the time slots for synchronization are naturally defined by mode-locked lasers exhibiting negligible timing jitter \cite{Kim2016AdvOptPhoton}.
State-of-the-art mode-locked lasers offer femtosecond pulse durations and sub-nanosecond repetition periods, which has a potential to enlarge the scale of the generated quantum state by orders of magnitude.
Using pulse-resolving homodyne detection, CV Einstein-Podolsky-Rosen pairs in a train of ultrashort pulses have been demonstrated on various kinds of platforms \cite{Okubo2008OL, shinjo2019pulse, Zhao2023OL}. However, time-domain multiplexed large-scale quantum state generation in a train of ultrashort pulses have not been demonstrated.

Here we report on the theoretical and experimental investigation of a time-domain multiplexing scheme for deterministically generating global multi-mode intensity squeezing in a train of ultrashort pulses.   
A OPA is used as a nonlinear beam-splitter to coherently combine delayed signal and idler pulses from another OPA, constituting a path unbalanced SU(1,1) interferometer (USUI) \cite{Yurke1986PRA, Ou2020APLPhotonic, Anderson2017Optica, Huo2022PRXQuantum,Zhang2023OptLett}.
Using pulse resolved detectors to perform joint measurement, we reveal the unique correlation structure of multi-mode state out of the USUI for the first time. 
On the one hand, each temporal mode is correlated with 5 other temporally adjacent modes.
On the other hand, 
the degree of multi-mode intensity squeezing depends not only on the gain of the OPAs, but also on the number of modes involved in the joint measurement, highlighting the global nature of the multimode squeezing.
Moreover, we experimentally demonstrate the generation of multi-mode intensity squeezing in a train of ultrashort pulses with temporal mode duration of about 10 ps. We characterize the multi-mode state when the number of modes involved in joint measurement is varied from 2–30 in different cases, and the results agree with theory. 
Our results extend time-domain multiplexed quantum state generation scheme to ultrashort pulses regime, and offers new perspectives on the correlation structure of state generated from an unbalanced SU(1,1) interferometer.

\section{\label{Theory}Experimental Principles}

\begin{figure}[htbp]
\centering
\includegraphics[width=0.8\textwidth]{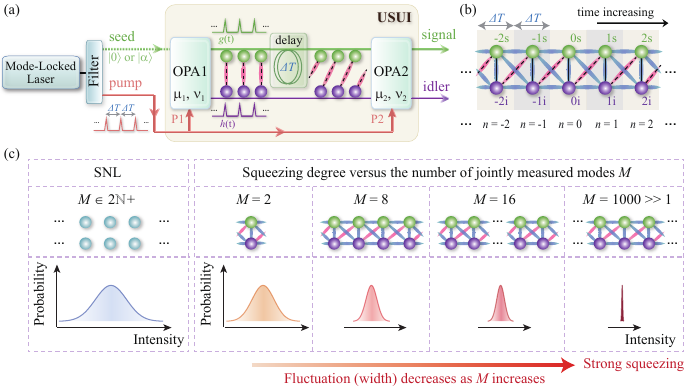}
\caption{\label{schematic} The schematics of generating global multimode state in a train of ultra-short pulses by time-domain multiplexing. (a) An unbalanced SU(1,1) interferometer (USUI) consisting of two pulse-pumped optical parametric amplifiers (OPA1, OPA2) with a delay line in between. The delay is introduced in the signal channel before coupling two outputs of OPA1 into OPA2, which functions as a nonlinear beam splitter. (b) Correlation structure of the multi-mode state. Besides the correlation (connect by thick pink lines) created by the parametric interaction in OPA1 (dashed black lines), five extra quantum correlation (thick blue lines) within each 4-mode unit, composed of the signal and idler modes in two adjacent time slots, are created by the interaction in OPA2 (solid black lines).
	(c) Global intensity squeezing out of the USUI operated in high gain regime. The simulations are conducted when $\mu_1=\mu_2=10$. Comparing with the corresponding shot noise level (SNL), the intensity noise fluctuation $R_d$ (see Eq.~\ref{Rd_M}) decrases with the increase of mode number $M$ involved in joint measurement. $R_d \to 0$ is obtained when $M\to \infty$. 
	$\Delta T$: the repetition period of pulsed pump; $g(t)$ (green) and $h(t)$ (purple): temporal function for signal and idler modes; 
	$n$: indices of time slots where temporal modes resident in; $s$ or $i$: mode indices for signal or idler. 
}
\end{figure}

The USUI consists of two identical OPAs (OPA1, OPA2) with a delay line in between, as shown in Fig.~\ref{schematic}(a). 
In each OPA, the pump (P1 and P2) out of a mode-locked laser propagates through the waveguide, where the optical parametric process takes place. 
Assuming that the $\chi^{(3)}$ nonlinearity is exploited, four wave mixing (FWM) occurs in the waveguide, 
in which two pump photons at $\omega_p$ are annihilated and a pair of signal and idler photons at $\omega_s$ and $\omega_i$ are created simultaneously.
Due to the energy and momentum conservation, we have $2\omega_p=\omega_s+\omega_i$.
Therefore, the signal and idler modes created by the same pump pulse are pulsed twin-beams possessing intensity difference squeezing between the two modes in the same time slot, but those created by different pump pulses are independent \cite{Zhao2023OL}. 
Assuming that the joint spectral function of the FWM parametric process is factorable \cite{Xueshi2015OE}, we consider signal and idler pulses, represented by the green and purple circles in Fig.~\ref{schematic}(a), with transform-limited temporal profiles  $g(t)$ and $h(t)$. Their corresponding spectral profiles are $\phi\left(\omega_{s} \right) = \frac{1}{\sqrt{2\pi}}\int g(t) e^{-i\omega_s t} dt$ and $\psi\left(\omega_{i} \right) = \frac{1}{\sqrt{2\pi}}\int h(t) e^{-i\omega_i t} dt$, respectively. 
Each temporal mode is labeled by a combination of two characters: one denotes the time slot indexed by $n\in\mathbb{Z}$, and the other denotes the signal ($s$) or idler ($i$) channel, as shown in Fig.~\ref{schematic}(b).
The signal and idler fields out of OPA1 can be written as \cite{Huo2022PRXQuantum}
\begin{eqnarray}
\hat{A}'_{ns} & = \mu_1 \hat A^{\mathrm{(in)}}_{ns}+ \nu_1 (\hat A_{ni}^{\mathrm{(in)}})^{\dagger} \nonumber\\
\hat{A}'_{ni} & = \mu_1  \hat A^{\mathrm{(in)}}_{ni}+ \nu_1  (\hat{A}_{ns}^{\mathrm{(in)}})^{\dagger},
\end{eqnarray}
where $\mu_1$ and $\nu_1$, satisfying $\mu_1^2 - \nu_1^2 =1$, are the amplitude gain of OPA1, and $\hat{A}^{\mathrm{(in)}}_{ns}  = \int \phi^{*}\left(\omega_{s} \right) \hat{a}_{s}\left(\omega_{s} \right) d \omega_{s}$ or $\hat{A}^{\mathrm{(in)}}_{ni} = \int \psi^{*}\left(\omega_{i} \right) \hat{a}_{i}\left(\omega_{i} \right) d \omega_{i}$ denotes the input of OPA1.

Before coupling the signal and idler fields into OPA2, which functions as a nonlinear beam splitter, the optical delay $\Delta T$ matching the repetition period of pump pulses is introduced in signal channel. 
The output of OPA2 can be written as \cite{Huo2022PRXQuantum} 
\begin{eqnarray}\label{OPA2-out}
\hat{A}_{ns}=&\mu_2 \mu_1 \hat A^{\mathrm{(in)}}_{ns} + \mu_2 \nu_1  (\hat A^{\mathrm{(in)}}_{ni})^\dagger + 
e^{i\theta} \nu_2 \mu_1  (\hat A^{\mathrm{(in)}}_{(n+1)i})^\dagger +  e^{i\theta} \nu_2 \nu_1 \hat A_{(n+1)s}^{(in)} \nonumber\\
\hat{A}_{ni}=&\mu_2 \mu_1 \hat A^{\mathrm{(in)}}_{ni} + \mu_2 \nu_1 (\hat A^{\mathrm{(in)}}_{ns})^\dagger + 
e^{i\theta} \nu_2 \mu_1  (\hat A^{\mathrm{(in)}}_{(n-1)s})^\dagger +  e^{i\theta} \nu_2 \nu_1 \hat A_{(n-1)i}^{(in)}
\end{eqnarray}
where $\theta$ is the phase difference between P1 and P2.  $\mu_2$ and $\nu_2$, satisfying $\mu_2^2 - \nu_2^2 =1$, are the gain of OPA2.
Eqs.~(\ref{OPA2-out}) shows the parametric interaction in OPA2 (black solid lines in Fig.~\ref{schematic}(b)) creates complex correlations that spread among adjacent time slots (thick blue lines in Fig.~\ref{schematic}(b)), resulting in a distinct multi-mode quantum correlation structure. 
For the sake of brevity, we consider $M$ ($M \in 2\mathbb{N}^{+}$) modes distributed over $M/2$ successive time slots, 
as illustrated in Fig.~\ref{schematic}(b) for the case of $M=10$. 
Successively applying Eqs.~(\ref{OPA2-out}) in time increasing order for all modes, we can obtain the covariance matrix formalism (see Appendix A for details), from which the features of the $M$-modes quantum state can be derived and characterized.

One way to characterize the correlation structure of the multi-mode state is analyzing the second-order intensity correlation functions in different cases. When the input of OPA1 is vacuum ($|0\rangle$), we have
\begin{equation}\label{eq:g_ij}
g^{(2)}(p,q) \equiv \frac{\langle\hat{A}_p^{\dagger}\hat{A}_q^{\dagger}\hat{A}_p\hat{A}_q\rangle}{\langle \hat{A}_p^{\dagger} \hat{A}_p \rangle \langle \hat{A}_q^{\dagger} \hat{A}_q\rangle}, 
\end{equation}
where $p$ or $q$ represents a specific mode index such as $0s$ or $1i$ (see Fig.~\ref{schematic}(b)). 
Without loss of generality, we take $p=0s$ as the reference mode and obtain the analytical expressions of $g^{(2)}(0s, q)$ by substituting the covariance matrix (in Appendix A) into the formalism derived in Ref. \cite{Vallone2019PRA} (See Appendix B for more details), as listed in Table 1. In the calculation, we assume the loss between OPA1 and OPA2 is neglectable.  
From Table 1, one sees that the USUI becomes equivalent to single OPA when the pump of OPA2 (OPA1) is turned off. In this condition, $\mu_2=1$ ($\mu_1=1$), we have $g^{(2)}(0s, -1i) = 1 + \mu_1^2/\nu_1^2$ ($g^{(2)}(0s, 0i)= 1 + \mu_2^2/\nu_2^2$) and $g^{(2)}(0s,q)=1$ for $q\neq -1i~(0i)$, indicating that the quantum correlation only exists between two modes of $0s$ and $-1i~(0i)$. To establish the correlations represented by the thick blue lines in Fig.~\ref{schematic}(b), both OPA1 and OPA2 should be operated in high gain regime and the gains of two OPAs should be balanced, i.e. $\mu_1=\mu_2=\mu\gg1$. In this case, we have $g^{(2)}(0s,0i) \to 2$, $g^{(2)}(0s,q) \to 1.25$ for $q\in\{-1i,-1s,1s,1i\}$ and $g^{(2)}(0s,q)=1$ for $q\notin\{0s, 0i, -1i,-1s,1s,1i\}$. Note we always have $g^{(2)}(0s,0s)\to2$ since each signal or idler pulse is in single temporal mode.
The results in Table I indicate that for any mode labeled by $ns(i)$ is correlated with 5 other modes in time slots of $n$, $n+1$ and $n-1$ (see Fig.~\ref{schematic}(b)).

\begin{table}[htpb]
\renewcommand{\arraystretch}{1.9}   
\centering
\label{tab_1}
\caption{Second-order intensity correlation $g^{(2)}(0s, q)$ between the reference mode $0s$ and mode $q$.}
\resizebox{\columnwidth}{!}{
	\begin{tabular}{|c|c|c|c|c|c|c|}
		\hline
		$g^{(2)}(0s, 0s)$ &$g^{(2)}(0s, -1i)$ & $g^{(2)}(0s, 0i)$ & $g^{(2)}(0s, -1s)$ & $g^{(2)}(0s, 1s)$ & $g^{(2)}(0s, 1i)$ & others  \\ 
		\hline
		2
		& $1+(\frac{\mu_1\nu_1\mu_2^2}{\mu_1^2\nu_2^2+\nu_1^2\mu_2^2})^2$         
		& $1+(\frac{\mu_2\nu_2(\mu_1^2+\nu_1^2)}{\mu_1^2\nu_2^2+\nu_1^2\mu_2^2})^2$     
		& $1+(\frac{\mu_1\nu_1\mu_2\nu_2}{\mu_1^2\nu_2^2+\nu_1^2\mu_2^2})^2$ 
		& $1+(\frac{\mu_1\nu_1\mu_2\nu_2}{\mu_1^2\nu_2^2+\nu_1^2\mu_2^2})^2$        
		& $1+(\frac{\mu_1\nu_1\nu_2^2}{\mu_1^2\nu_2^2+\nu_1^2\mu_2^2})^2$
		& 1 \\ 
		\hline
	\end{tabular}
}
\end{table}

The other way to characterize the correlation structure is analyzing the intensity noise among multiple modes.
We consider the linear combination of photon-number $\hat N_d$ among $M$ ($M \in 2\mathbb{N}^{+}$) modes
\begin{equation}\label{N_d}
\hat N_d =  \sum_{n = 1}^{M/2} \hat N_{ns} - \sum_{n = 1}^{M/2} \hat N_{(n+m)i} ,
\end{equation}
where $\hat N_{ns} = \hat A^\dagger_{ns}\hat A_{ns}$ and $\hat N_{(n+m)i}=\hat A^\dagger_{(n+m)i}\hat A_{(n+m)i}$ are the photon number operators for the pulses in signal and idler channel, and $m \in \mathbb{N}$ refers to the time delay relative to the $n$-th slot. The characterization of $\hat N_d$ requires to jointly measure the intensities among $M$ modes when two OPAs are operated in high gain regime. To implement the measurement, we exploit the self-homodyne detection by injecting the seed from the signal input port of OPA1 \cite{d1998self, Yan2025OL}. 
The seed in coherent state ($| \alpha \rangle \gg 1$) is synchronized with pump pulses, and its power is orders of magnitude lower than that of pump. 
Under this condition, the intensity fluctuation of any individual pulse mode is $ {(\langle \hat N_q^2\rangle-\langle \hat N_q\rangle^2)}/{\langle \hat N_q\rangle}  \approx  4\mu^4$, which is $4\mu^4$ times high than that of shot noise level (SNL). We find the intensity squeezing of $\hat N_d$ is optimized when $m=0$, which means the signal and idler pulses involved in the joint measurement should be in same time slots (see Appendix C for details). In this case, we have the intensity noise  normalized to SNL:  
\begin{eqnarray}
\label{Rd_M}
R_d^{(M)}  \equiv \frac{\langle \hat N_d^2 \rangle - \langle \hat N_d \rangle^2 }
{\sum_{n=1}^{M/2} (\langle \hat N_{ns} \rangle + \langle \hat N_{ni} \rangle)} \approx \frac{1}{M} + \frac{1}{8\mu^4}.
\end{eqnarray}
where $\mu^2$ refers to the power gain of both OPA1 and OPA2. 

Eq.~(\ref{Rd_M}) shows that $R_d^{(M)} \to 0$ is achievable when both the number of modes $M$ involved in the joint measurement and the power gain of OPA $\mu^2$ are infinitely large. 
As illustrated in Fig.~\ref{schematic}(c), even in the high-gain regime ($\mu_1=\mu_2=\mu=10$), the intensity distribution retains a finite width for $M=2$ (or $M=8$), corresponding to a normalized intensity noise of $R_d\approx0.5~(0.125)$. Since $R_d$ is defined as the variance of the intensity normalized to the SNL, the width of the intensity distribution scales with $\sqrt{R_d}$. Consequently, the intensity distribution keeps narrowing as the number of jointly measured modes $M$ increases. Only in the limit of a sufficiently large $M$ ($M\to\infty$) can the intensity noise fluctuation approach zero, i.e., $R_d\to 0$.
Therefore, the analysis indicates the structure of the global multi-mode squeezing produced by the USUI is fundamentally different from that obtained by other approaches, in which the nullifiers are achievable within the joint measurement of 4 modes \cite{Yokoyama2013}.
Moreover, Eq.~(\ref{Rd_M}) indicates that to achieve optimal intensity squeezing at a given power gain $\mu^2$, the number of modes included in the joint measurement must increase with $\mu^2$; specifically, reaching the optimal squeezing requires at least $M \sim 8 \mu^4$ modes.
In contrast, individual OPA with power gain of $4\mu^4$ (with $\mu^2\gg1$) equivalent to a balanced SU(1,1) interferometer only can generate intensity squeezing between two modes (or intensity difference squeezing) $R_d^{(2)}=\frac{1}{8\mu^4}$ \cite{Lou2023SciChina}, where the squeezing degree only depends on the gain of OPA.

\section{\label{sec:Experiment}Experimental Results}
\begin{figure}[htpb]
\centering
\includegraphics[width=0.6\textwidth]{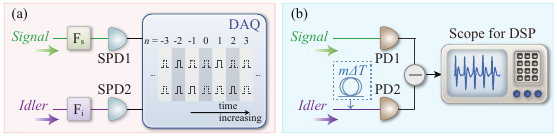}
\caption{\label{setup}The detection schemes for
(a) characterizing the structure of multi-mode state by intensity correlation $g^{(2)}$ between modes $0s$ and $q$ and for (b) measuring the multi-mode intensity squeezing by pulse resolving self-homodyne and digital signal processing (DSP).	
Dashed-box: optional delay line. $\mathrm{F_{s(i)}}$: optical filters; SPD, single photon detector; DAQ, data acquisition system; PD, photo diode.
}
\end{figure}
The detailed experimental schematic of Fig.~\ref{schematic}(a) is described in Appendix D.  The nonlinear medium of each OPA is a piece of dispersion-shifted fiber (DSF) with length and zero-dispersion wave length of 300 m and 1548.5 nm, respectively. The central wavelength of pump is 1549.3 nm.
The DSFs are immersed in liquid nitrogen (77 K) to suppress Raman noise.
Under these conditions, the phase matching condition of FWM in each OPA is satisfied \cite{Huo2022PRXQuantum}. The pump and seed are obtained by carving the output of a mode-locked laser with the central wavelength, 5 dB bandwidth and repetition period $\Delta T$ of 1550 nm, 40 nm, and 20 ns, respectively \cite{Zhao2023OL}. The durations of the transform limited pump and seed pulses are about 9 ps and 5 ps, respectively. To ensure of gain of each OPA is high enough, the pump is amplified by using erbium-doped fiber amplifier. When the average power of pump (P1 and P2) is 1.8 mW and the mode matching between the seed and pump in all degree of freedoms is optimized, the power gain ($\mu^2$) of each OPA is about 38. The gain bandwidth (about 40 nm) of each OPA is broad, the central wavelengths of signal and idler fields are chosen to be 1533.0 nm and 1566.0 nm, respectively. Moreover, during the measurement of intensity correlation function $g^{(2)}$, the input of OPA1 is in vacuum, and we send the output of USUI through a dual-band filter (Fig.~\ref{setup}(a)). In this case, the signal and idler fields with 3 dB bandwidth of about 0.6 nm is nearly in spectral factorable state. 

\begin{figure}[htpb]
\centering
\includegraphics[width=0.7\textwidth]{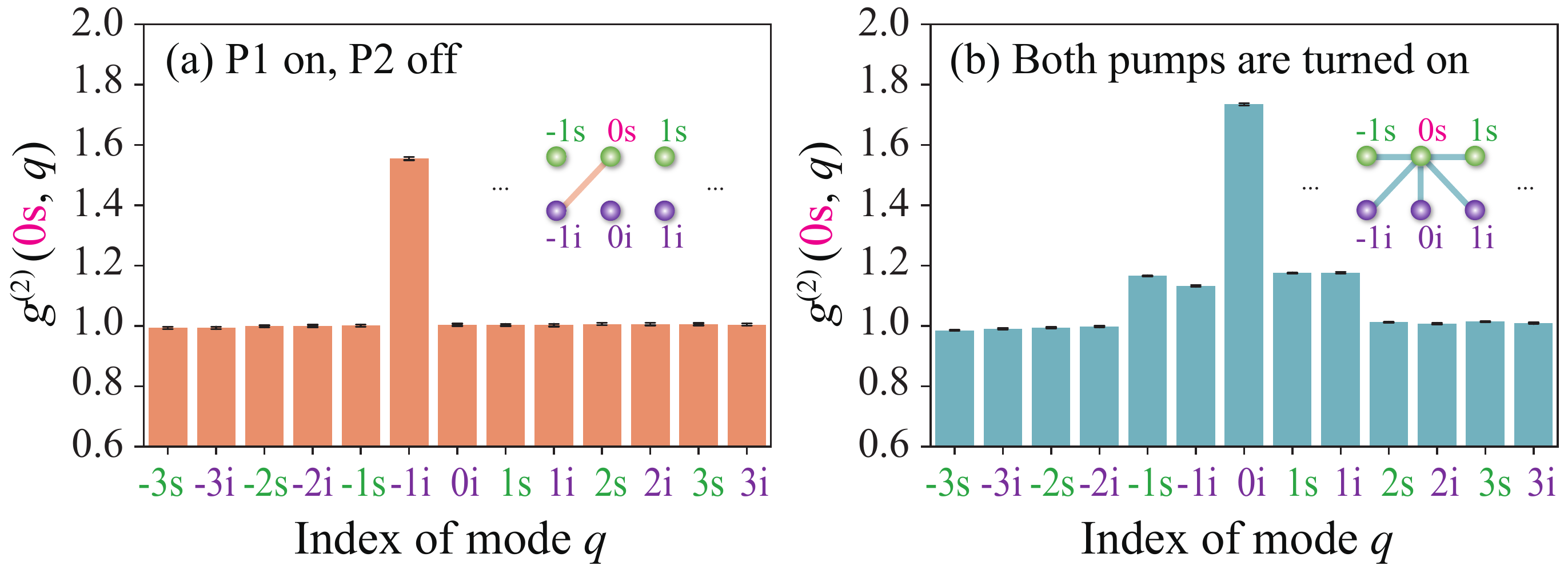}
\caption{\label{g_ij_bar} Measured intensity correlation $g^{(2)}(0s, q)$ between the reference mode $0s$ and another $q$ for
(a) $\mu_1^2\approx 10,\mu_2^2=1$ and (b) $\mu_1^2\approx\mu_2^2\approx 10$. 	
}
\end{figure}

We begin with measuring intensity correlation function $g^{(2)}$ between the modes $0s$ and $q$.
As shown in Fig.~\ref{setup}(a), the signal and idler fields are respectively measured by two single photon detectors (SPDs). 
Each SPD with response time of about 200 ps can resolve photons from different pulses. 
Since the measurement of $g^{(2)}$ is irrelevant to detection efficiency, an attenuator is placed infront of each SPD to ensure the detection rate of each SPD is less than 0.1 photon/pulse. Then $g^{(2)}(0s, q)$ can be extracted from the ratio of coincidence rate between $0s$ and $q$ to their accidental coincidence rate. 
We first measure $g^{(2)}$ when P2 is off, and the power gain of OPA1 is $\mu_1^2\approx10$. As shown in Fig.~\ref{g_ij_bar}(a), only $g^{(2)}(0s, -1i)\approx 1.56$ is obviously larger than 1 and all the other cases satisfies $g^{(2)}=1$ within error bars. Then, we turn on both pumps and repeat the measurement when $\mu_1^2\approx\mu_2^2\approx\mu^2\approx 10$.
As shown in Fig.~\ref{g_ij_bar}(b), for all $g^{(2)}$ involving the reference mode $0s$, correlations with $g^{(2)} > 1$ are observed for the mode pairs $(0s,-1s)$, $(0s,-1i)$, $(0s,0i)$, $(0s,1s)$, and $(0s,1i)$, with values of approximately $1.17$,  $1.13$,  $1.73$,  $1.18$, and  $1.18$, respectively, indicating the presence of correlations between these modes.
We also measure the second-order auto-correlation of signal and idler field in this condition, and the result are $g^{(2)}(0s,0s) \approx g^{(2)}(0i,0i) \approx 1.93$, showing the joint spectral function of the OPAs are near factorable.  
The results in Fig.~\ref{g_ij_bar}(b) verifies the correlation structure in Fig.~\ref{schematic}(b). The measured $g^{(2)}$ between $0s$ and $q$ modes is slightly lower than that prediction in Table 1, which predict $g^{(2)}(0s,0i)\to 2$, $g^{(2)}(0s, \pm1s)\to 1.25$, and $g^{(2)}(0s, \pm1i)\to 1.25$ when $\mu^2 = 10$. 
We think this discrepancy is mainly due to the reduction of collection efficiency of twin photons induced by the usage of narrowband filters.

Next, we measure the multi-mode intensity squeezing by using the pulse resolvable self-homodyne detection (see Fig.~\ref{setup}(b)). In this case, weak signal with average power of about 0.13 $\mathrm{\mu W}$ functioning as the seed is injected into OPA1, and the USUI is locked by actively controlling the phase so that the intensities of signal and idler fields generated from the amplification of seed are maximum \cite{WangHailong2015APL}. The signal and idler fields are then respectively sent to two photo diodes (PD1 and PD2) with bandwidth of 100 MHz, which form a balanced detection being able to resolve individual pulses. In order to analyze intensity noise for signal and idler fields in different time slots, an optional delay line used to introduce delay time $m\Delta T$ is placed in front of PD2. The output current of balanced detector is recorded by a digital oscilloscope for digital signal processing. The detection efficiency of system (including the quantum efficiencies of PD $\sim 81\%$ and transmission efficiency of optical components $\sim 78\%$) is $\sim 63\%$. For the intensity noise among $M$ modes, the mode number of individual PD1 and PD2 involved in the joint measurement is $M/2$.
For each single data acquisition process, $25\times 10^6$ samples are recorded at a sampling rate of $5\times 10^9$ sample/s,  corresponding to a 5 ms time window containing 250,000 optical pulses. We integrate every $M/2$ consecutive electronic pulses output from the balanced detector, yielding $\sim \frac{250000}{(M/2)}$ integrated values. The intensity noise is then characterized by the variance of these integrated values, and
normalized to the corresponding SNLs, which are carefully calibrated using the coherent states with powers equal to those of the signal and idler fields (see Appendix E for the details of SNL calibration and data acquisition process).

\begin{figure}[htpb]
\centering
\includegraphics[width=0.7\textwidth]{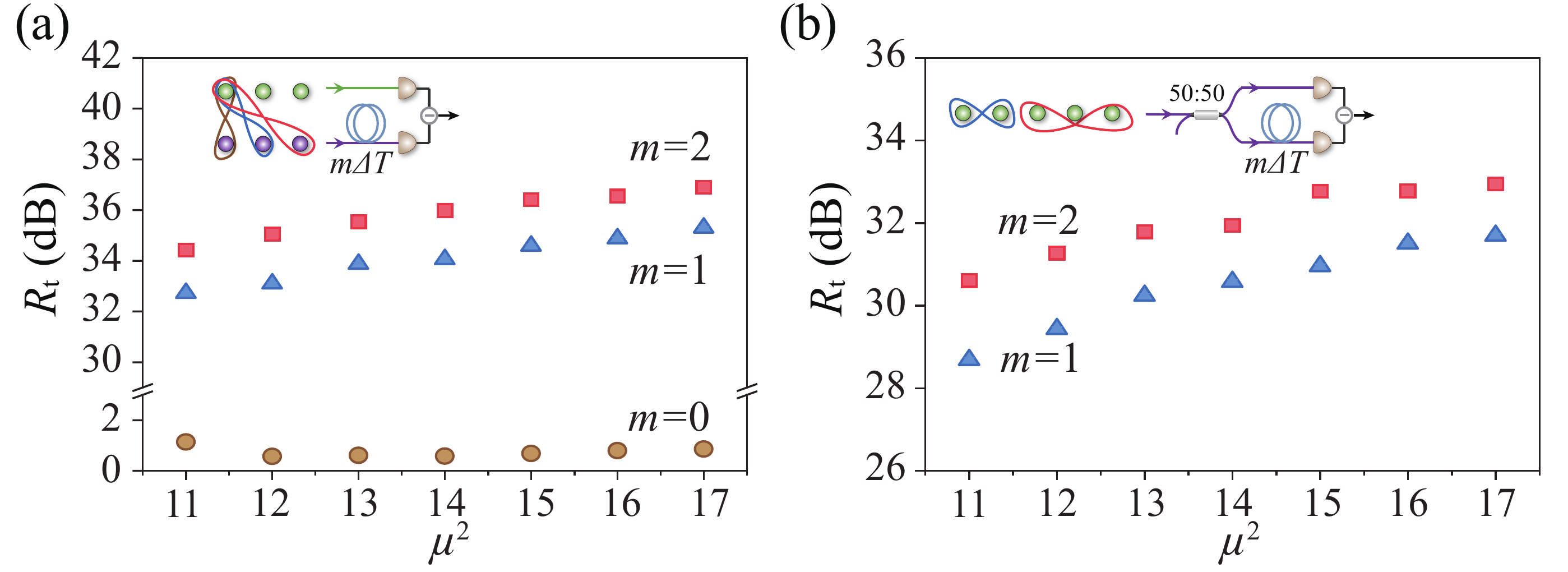}
\caption{\label{2_mode_IDS} Normalized two-mode intensity noise $R_t$ versus the power gain $\mu^2$ of OPA for (a) signal and idler modes when the interval $m\Delta T$ between their time slots is $m=0,1,2$ and (b) two signal modes in time slots with $m=1,2$.   
}
\end{figure}

Figure~\ref{2_mode_IDS} shows the normalized noise $R_t=(\langle\hat{N}_t^2\rangle-\langle\hat{N}_t\rangle^2)/(\langle \hat{N}_p \rangle+\langle \hat{N}_q \rangle)$ of the two-mode intensity difference between modes $p$ and $q$ as a function of the power gain $\mu^2$ of the OPAs for different cases, where $\hat{N}_t=\hat{N}_p-\hat{N}_q$. 
The relative delay $m\Delta T$ between the modes respectively measured by PD1 and PD2 can be changed by setting $m=0, 1, 2$ (see the insets). 
Fig.~\ref{2_mode_IDS}(a) plots the data when the two modes are in signal and idler channel respectively. 
It is clear that the variation tendencies of the three cases are different. When $m=2$ and $m=1$, the noise increases with $\mu^2$ and is more than 30 dB higher than the SNL in high gain regime. The data for $m=1$ is slightly lower than that for $m=2$, because two modes in time slots with $m=2$ are independent while two modes in neighboring slots with $m=1$ are correlated to some extent (see Fig.~\ref{schematic}(b)). In contrast, in the case of $m=0$, the noise decreases with the increase of $\mu^2$ when $\mu^2<12$, and the noise level remains close to SNL for $\mu^2>12$. Notably, the data for $m=0$ is significantly lower than the other two cases, indicating the correlation between the same time slot is the strongest. We also measure $R_t$ when two modes are from individual signal (idler) channel, which is realized by passing signal (idler) field through a 50/50 beam splitter (BS) and launching two outputs of BS to PD1 and PD2 (see the inset in Fig.~\ref{2_mode_IDS}(b)), respectively.  As shown in Fig.~\ref{2_mode_IDS}(b), the results for the cases of $m=1$ and $m=2$ are similar to those in Fig.~\ref{2_mode_IDS}(a), except that the noise levels are lower than that in Fig.~\ref{2_mode_IDS}(a) by about 3 dB due to the vacuum noise induced by 50/50 BS.
We also measure the $m=0$ case of Fig.~\ref{2_mode_IDS}(b), and the measured noise remains close to the SNL because the common-mode fluctuations are largely canceled by the balanced detection.  
The results in Fig.~\ref{2_mode_IDS} being consistent with those in Fig.~\ref{g_ij_bar} agree with the theory predictions. 
Notice that the noise level for the case of $m=0$ in Fig.~\ref{2_mode_IDS}(a) has not reached the theory prediction of 3 dB below SNL, we think the main reason is due to the extra noise of seed injection \cite{Yan2025OL}.

\begin{figure}[htpb]
\centering
\includegraphics[width=0.7\textwidth]{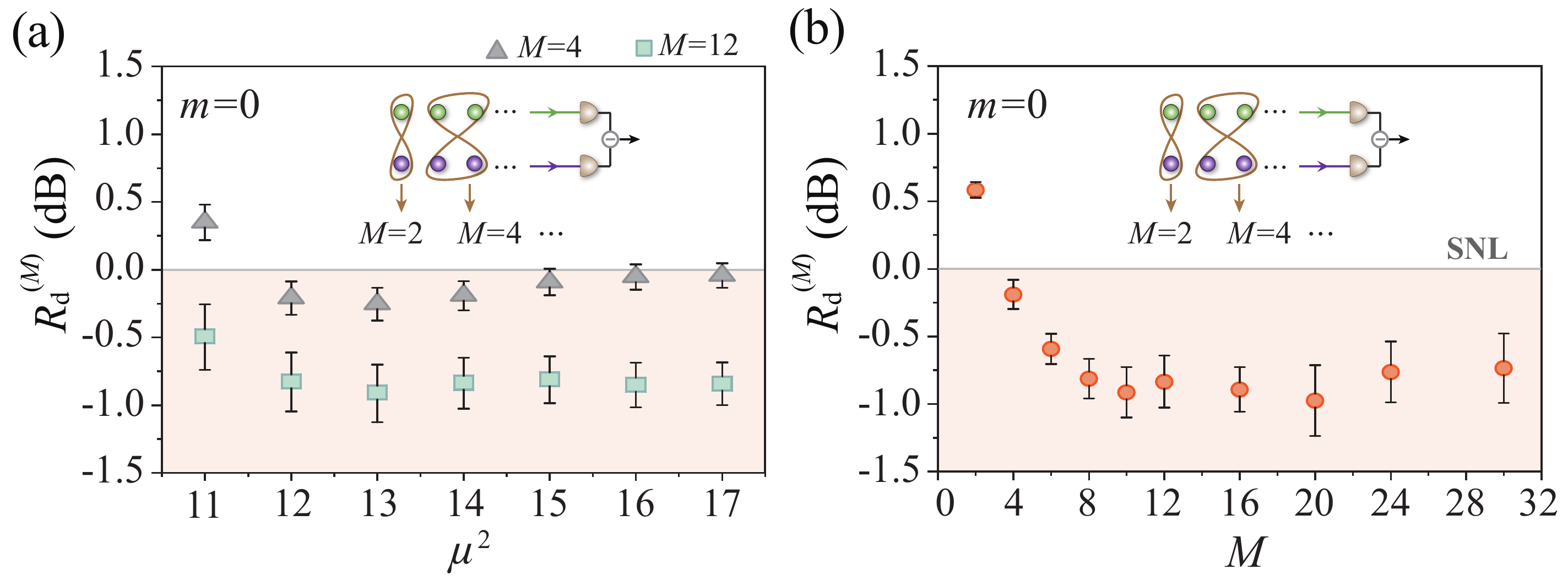}
\caption{\label{M_mode_IDS} 
Normalized intensity noise $R_d^{(M)}$ measured by varying the gain of OPA and mode number $M$ involved in joint measurement. (a) $R_d^{(M)}$ versus the power gain $\mu^2$ when $M=4, 12$; (b) $R_d^{(M)}$ versus $M$ when $\mu^2\approx14$.  
}
\end{figure}

Figure~\ref{M_mode_IDS} demonstrates the normalized intensity noise $R_d^{(M)}$ when the mode number $M$ of joint measurement is greater than 2. In the experiment, the successive time slots occupied by the signal and idler pulses are the same $(m=0)$. Fig.~\ref{M_mode_IDS}(a) shows $R_d^{(M)}$ versus the power gain $\mu^2$ when $M=4, 12$. For $R_d^{(M)}$ measured under a fixed gain, $R_d^{(12)}$ is obviously lower than $R_d^{(4)}$. 
Moreover, the lowest value of $R_d^{(4)}\approx- 0.25$ dB is below the SNL and $R_d^{(4)}$ stays lower than SNL for $\mu^2\ge12$, indicating $R_d^{(4)}$ is lower than $R_d^{(2)}$. Here, $R_d^{(2)}$ is the same as the two-mode intensity noise $R_t$ between the signal and idler modes for $m=0$ (see the brown circles in Fig.~\ref{2_mode_IDS}(a)).
The results manifest that involving more modes in the joint measurement in a key to extract out the quantum correlation. To further verify this point, we measure $R_d^{(M)}$ by fixing $\mu^2\approx14$ and varying $M$ from 2 to 30. As shown in Fig.~\ref{M_mode_IDS}(b), $R_d^{(M)}$ rapidly decrease with the increase of $M$ when $M<10$, and stays at around -0.9 dB for $M$ in the range of about 10--20. When $M>20$, $R_d^{(M)}$ shows an upward trend due to the influence of extra noise but remains below the SNL. Note that the intensity squeezing in Fig.~\ref{M_mode_IDS} agreeing with the prediction of Eq.~(\ref{Rd_M}) is currently confined by detection efficiency and extra noise of the seed injection. Once these technique issues are solved, involving more modes with number $M>30$ will be necessary for efficiently utilizing the multimode quantum correlations and for measuring the improved intensity squeezing.

\section{Summary and Discussion}
In conclusion, we have successfully demonstrated the multi-mode continuous variable quantum state by using the USUI with mode-locked laser as pump, in which the time-domain multiplexing is realized by following one high gain OPA with the optical delay and another OPA functioning as a nonlinear beam splitter. Exploiting the pulse resolved intensity detection, 
we find that the degree of multi-mode intensity 
squeezing is not solely determined by the gain of OPAs. 
Even if the gains of OPAs are infinitely large and the detection efficiency is perfect, the fluctuation of intensity noise approaching to zero is still unattainable unless the number of modes involved in the joint measurement of intensities is also infinitely large. 
The state is fundamentally different from that generated by the time-domain multiplexing realized by delay and linear beam splitter \cite{Yokoyama2013}, in which the nullifiers is achievable when the mode number of joint measurement is 4.
Moreover, we have experimentally generated the global multi-mode intensity squeezing in a train of ultrashort pulses. 
For the number of jointly measured modes up to $M=30$, we observe that the degree of multimode intensity squeezing increases with $M$, and the squeezing degree is $\sim 0.9$ dB lower than the SNL for $M>10$. Limited by the technical imperfections, the experimentally measured correlation degrees are lower than the theorical expectations, but the variation tendencies of experimental results agree with theory predictions. 
In our experiment, the repetition period and duration of temporal mode are 20 ns and $\sim 10$ ps, respectively, which establish new records for generating quantum state by the configuration of time-domain multiplexing. 
By exploiting the state-of-the-art high repetition rate mode-locked laser and photo-detectors with GHz bandwidth, the scale of the quantum state can be further enlarge to more than $10^9$ modes per second. 
The present work demonstrates global multimode intensity squeezing, whereas our preliminary theoretical analysis suggests that similar global behavior can also emerge in quadrature measurements~\cite{guo2023multimode_arxiv}.
Our investigations indicate that the USUI is a promising platform for generating global multimode quantum state, which will be suitable for multi-user quantum networks~\cite{JiaXiaojun2024PhysRevRes,wu2020passive}.

The observation of global multimode squeezing greatly depends on the distinct coherence feature possessed by the emission field from USUI. It is well known that for the thermal light occupying broadband continuum spectrum, we have $\tau_c\propto\frac{1}{\Delta\nu}$ and $g^{(2)}(\tau)=1+|g^{(1)}(\tau)|^2$, where $\tau_c$ and $\Delta\nu$ denote the coherence time and frequency bandwidth, and $|g^{(1)}(\tau)|=e^{-\tau^2/\tau_c^2}$ is the first order  coherence degree. Traditionally, we have $g^{(1)}(\tau)\to 0$ and $g^{(2)}(\tau)\to 1$ when $|\tau|\gg\tau_c$.  However, in our experiment, we obtain $g^{(2)}(\tau)\approx1.17>1$ when $\tau=\pm20$~ns is orders of magnitudes larger than the coherence time $\tau_c\approx 10$~ps. 
On the other hand, for the signal and idler thermal fields created by traditional parametric process with pulsed laser serving as the pump, the intensity correlation between signal and idler fields $g_{si}^{(2)}>1$ is observable within one time window defined by pump pulse duration~\cite{MandelWolf1995}. However, in our experiment, $g_{si}^{(2)}>1$ is simultaneously observable in three different time windows with separation several orders magnitudes larger than pump pulse duration. This is because the light emission mechanism of high gain USUI is novel. In the sense that traditional optical parametric amplification falls into two categories: phase-sensitive and phase-insensitive amplification, the high gain USUI is unclassifiable under existing regulations. Moreover, there are plenty of space for tailoring the coherence of the light out of USUI by flexibly constructing the delay line with different kinds of length combination. We believe the high gain USUI provides us with a new tool for engineering the coherence of ultrafast thermal light, which will be valuable for precision measurement and for the application scenarios suffered from the speckle of lasers. 

\begin{acknowledgments}
	We would like to thank Prof. Qiongyi He for fruitful discussion. This work was supported by the Quantum Science and Technology - National Science and Technology Major Project (Nos. 2024ZD0302403 and 2024ZD0300804), the National Natural Science Foundation of China (Nos. 12461160325 and 12504421), and the China Postdoctoral Science Foundation (No. 2024M762347).
\end{acknowledgments}

\section*{Data Availability}
The data that support the findings of this article are not publicly available. More information and data are available upon reasonable request from the authors.

\appendix

\renewcommand{\theequation}{A\arabic{equation}}
\setcounter{equation}{0} 

\renewcommand{\thefigure}{A\arabic{figure}} 
\setcounter{figure}{0}                       

\section{Derivation of the covariance matrix in complex basis} \label{Sec1}
\begin{figure}[h]
	\centering
	\includegraphics[width=0.5\textwidth]{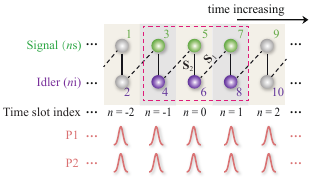}
	\caption{\label{fig_state} Nonlinear interactions among the output modes of the USUI. Dashed or solid black lines: parametric processes from OPA1 or OPA2 between connected modes. P1 and P2 denote the pump pulses of OPA1 and OPA2, respectively.
	}
\end{figure}

We consider a multi-mode quantum state out of an unbalanced SU(1,1) interferometer (USUI) consisting of two optical parametric amplifiers (OPAs).
Each OPA is pumped by a train of ultra-short pulses filtered from a mode-locked laser.
The time slots of the multi-mode state are defined by the repetition rate of the pump pulses, whose timing jitter is negligible.
Each pulse-defined time slot contains two frequency-distinct signal and idler channels, as shown in Fig.~\ref{fig_state}; hence,
$M/2$ time slots corresponds to a total of $M$ modes, where $M \in 2\mathbb{N}^+$ ( $\mathbb{N}^+ = {1,2,3,\dots}$).
Within the Gaussian quantum state regime \cite{Adesso2014OpenSys}, such an $M$-mode quantum state is fully characterized by its displacement vector and covariance matrix. 
Since pump pulses for both OPAs always exist before and after the analyzed $M/2$ time slots, the boundary modes can interact with modes outside the $M$-mode system. 
To account for this effect, we include two additional time slots before and after the analyzed window and consider an extended system with $M+4$ modes.
The displacement vector and covariance matrix of the $M$-mode quantum state are then obtained by tracing out the 4 additional boundary modes. 

Without loss of generality, we derive the covariance matrix for the output state of an USUI
with $M=6$ (see Fig.~1(b) of the main text, from $n=-1$ to $1$). 
We do so by considering an extended system with $M=10$ (see Fig.~\ref{fig_state}).
We note this approach can be directly extended to $M \to \infty$.
To ensure compatibility with matrix notation, as illustrated in Fig.~\ref{fig_state}, we redefine each mode index with a positive integer. In complex basis, the displacement vector $\mathbf{d'}^{(10)}$ can be written as \cite{Adesso2014OpenSys}
\begin{align} 
\label{d}
\mathbf{d'}^{(10)} =& \left[\begin{matrix} \bm{\alpha} &  \bm{\alpha^*}  \end{matrix}\right]^T, 
\end{align}
where $\bm{\alpha} = [ \langle{\hat A_1}\rangle, \langle{\hat A_2}\rangle, ..., \langle{\hat A_{10}}\rangle  ]$ and $\bm{\alpha^*} = [ \langle{\hat A_1^{\dagger}}\rangle, \langle{\hat A_2^{\dagger}}\rangle, ..., \langle{\hat A^{\dagger}_{10}}\rangle  ]$. The prime sign indicates the quantum system with 4 boundary modes
, and $\hat A_1,\cdots ,\hat A_{10}$ are annihilation operators of the corresponding modes. 
We note that, for the spontaneous parametric process in OPA1, the input to OPA1 is viewed as vacuum, and the displacement vector $\mathbf{d'}^{(10)}$ reduces to zero. 
In contrast, 
coherent seeding to OPA1 will introduce displacement to $\mathbf{d'}^{(10)}$ but dose not change the covariance matrix.

The covariance matrix in complex basis $\sigma'^{(10)}$ can be written as
\begin{equation}
\label{CM_def}
\sigma'^{(10)} = \left[\begin{matrix} \mathbf{A'}^{(10)} & \mathbf{B'}^{(10)} \\ (\mathbf{B'}^{(10)})^* & (\mathbf{A'}^{(10)})^* \end{matrix}\right],
\end{equation} 
where the element in the $p$-th row and $q$-th column ($p,q$ are integers and $p,q \in [1,10]$) of the submatrices $\mathbf{A'}^{(10)}$ and $\mathbf{B'}^{(10)}$ are given by
\begin{eqnarray}
A_{p,q}'^{(10)}  =& \langle \hat A_p \hat A^\dagger_q \rangle + \langle \hat A^\dagger_q \hat A_p  \rangle - 2 \langle{\hat A_p}\rangle \langle{\hat A_q^\dagger}\rangle, \nonumber\\
B_{p,q}'^{(10)} =& 2 \langle \hat A_p \hat A_q \rangle  - 2 \langle{\hat A_p}\rangle \langle{\hat A_q}\rangle.
\end{eqnarray}

The non-dengereate parametric interaction in an OPA can be modeled by symplectic transformation for two-mode squeezing $\mathbf{S}_{T}$, which has the form of:
\begin{equation}
\label{eq:stable_OPA}
\left[ \begin{matrix} \hat{A}_{p}  & \hat{A}_{q} & \hat{A}^\dagger_{p} & \hat{A}^\dagger_{q} \end{matrix} \right]^\mathbf{T} 
\mapsto \mathbf{S}_{T} \left[ \begin{matrix} \hat{A}_{p}  & \hat{A}_{q} & \hat{A}^\dagger_{p}  & \hat{A}^\dagger_{q}\end{matrix} \right] ^\mathbf{T},
\end{equation}
where 
\begin{equation}
\label{eq:S_tmsq}
\mathbf{S}_{T} = \left[\begin{matrix} 
	\mu               &  0                 &  0               & e^{i\varphi} \nu   \\
	0              &  \mu               & e^{i\varphi}\nu   & 0                 \\
	0              &  e^{-i\varphi} \nu  & \mu              & 0                 \\
	e^{-i\varphi}\nu  &  0                 & 0                & \mu   
\end{matrix}\right],
\end{equation}
$\mu$ and $\nu$ are the parametric gain of the OPA, satisfying $\mu^2-\nu^2=1$.
The squeezing angle $\varphi$ is determined by the pump phase of the OPA.
%
The symplectic transformation matrix of two-mode squeezing on mode pair $p$ and $q$ for $M$-mode system $\mathbf{S}_{T}^{(M,p,q)}$ can be obtained by generalizing the two-mode transformation $\mathbf{S}_{T}$ in Eq.~(\ref{eq:S_tmsq}), and takes the following block matrix form 
\begin{equation}
\label{S_tmsq}
\mathbf{S}_{T}^{(M,p,q)} = \left[\begin{matrix} \mathbf{S}_{A} & \mathbf{S}_{B} \\ \mathbf{ S}^*_{B} & \mathbf{ S}^*_{A} \end{matrix}\right],
\end{equation}
where $\mathbf{S}_{A}$ is a diagonal matrix satisfying $S_{A}(p,p) = S_{A}(q,q)=\mu$ and all other diagonal elements are equal to 1; matrix $\mathbf{S}_{B}$ only has two non-zero term satisfying $S_{B}(p,q) = S_{B}(q,p)=e^{i\varphi} \nu$ and all the other elements are zero.

We denote the parametric gains of OPA1(2) in our state generation scheme by $\mu_{1(2)}$ and $\nu_{1(2)}$. 
OPA1 operates in a phase-insensitive regime for both vacuum and coherent-state inputs at the signal channel. Therefore, without loss of generality, we set $\varphi = 0$ for the parametric interaction in OPA1. 
For OPA2, we introduce a single parameter $\theta$ to characterize its squeezing angle. 
By substituting $\mu_{1(2)}$, $\nu_{1(2)}$, and $\theta$ into Eq.~(\ref{S_tmsq}), 
the corresponding symplectic transformation matrices describing the parametric interactions in OPA1 and OPA2 can be readily obtained, which we denote as $\mathbf{S}_{1}^{(M,p,q)}$ and $\mathbf{S}_{2}^{(M,p,q)}$, respectively. 
The transformation associated with each parametric interaction can be formulated as
\begin{eqnarray}
	\label{state_evo}
	&\sigma'^{(10)}   \mapsto \mathbf{S}_{1(2)}^{(10,p,q)} \sigma'^{(10)} (\mathbf{S}_{1(2)}^{(10,p,q)})^\dagger, 
\end{eqnarray}
For our state generation scheme, the temporal order of the parametric interactions follows two rules: (1) Operations labeled by a larger time slot index $n$ always happen after those with a smaller time slot index; (2) For two operations connected to a given idler mode, the squeezing operation from OPA1 always precedes that from OPA2. Specific to the case of Fig.~\ref{fig_state}, 
	9 squeezing operations are applied sequentially to the mode pairs $\{2,3\}$, $\{2,1\}$, $\{4,5\}$, $\{4,3\}$, $\{6,7\}$, $\{6,5\}$, $\{8,9\}$, $\{8,7\}$, and $\{9,10\}$. Among them, 4 operations originate from OPA1 (dashed lines in Fig.~\ref{fig_state}), while the remaining 5 are from OPA2 (solid lines in Fig.~\ref{fig_state}). 
For the simplicity of expression, we introduce notations $V_{1(2)} =\mu_{1(2)}^2 +\nu_{1(2)}^2 $ and $c_{1(2)} = \mu_{1(2)}\nu_{1(2)}$, and the block matrices $\mathbf{A'}^{(10)}$ and $\mathbf{B'}^{(10)}$ in Eq.~(\ref{CM_def}) read as follows:
	\begin{equation}
		\label{CM_A10}
		\mathbf{A'}^{(10)} = \left[\begin{matrix} A'_{1,1} & 0 & 2 c_{1} c_{2} e^{i \theta} & 0 & 0 & 0 & 0 & 0 & 0 & 0\\0 & A'_{2,2} & 0 & 2 c_{1} c_{2} e^{- i \theta} & 0 & 0 & 0 & 0 & 0 & 0\\2 c_{1} c_{2} e^{- i \theta} & 0 & V_{1} V_{2} & 0 & 2 c_{1} c_{2} e^{i \theta} & 0 & 0 & 0 & 0 & 0\\0 & 2 c_{1} c_{2} e^{i \theta} & 0 & V_{1} V_{2} & 0 & 2 c_{1} c_{2} e^{- i \theta} & 0 & 0 & 0 & 0\\0 & 0 & 2 c_{1} c_{2} e^{- i \theta} & 0 & V_{1} V_{2} & 0 & 2 c_{1} c_{2} e^{i \theta} & 0 & 0 & 0\\0 & 0 & 0 & 2 c_{1} c_{2} e^{i \theta} & 0 & V_{1} V_{2} & 0 & 2 c_{1} c_{2} e^{- i \theta} & 0 & 0\\0 & 0 & 0 & 0 & 2 c_{1} c_{2} e^{- i \theta} & 0 & V_{1} V_{2} & 0 & 2 c_{1} c_{2} e^{i \theta} & 0\\0 & 0 & 0 & 0 & 0 & 2 c_{1} c_{2} e^{i \theta} & 0 & V_{1} V_{2} & 0 & 2 c_{1} c_{2} e^{- i \theta}\\0 & 0 & 0 & 0 & 0 & 0 & 2 c_{1} c_{2} e^{- i \theta} & 0 & A'_{9,9} & 0\\0 & 0 & 0 & 0 & 0 & 0 & 0 & 2 c_{1} c_{2} e^{i \theta} & 0 & A'_{10,10} \end{matrix}\right]
\end{equation}
and
	\begin{equation}
		\label{CM_B10}
		\mathbf{B'}^{(10)} = \left[\begin{matrix}0 & B'_{1,2} & 0 & 2 c_{1} \nu_{2}^{2} e^{2 i \theta} & 0 & 0 & 0 & 0 & 0 & 0\\B'_{2,1} & 0 & 2 c_{1} \mu_{2}^{2} & 0 & 0 & 0 & 0 & 0 & 0 & 0\\0 & 2 c_{1} \mu_{2}^{2} & 0 & 2 V_{1} c_{2} e^{i \theta} & 0 & 2 c_{1} \nu_{2}^{2} e^{2 i \theta} & 0 & 0 & 0 & 0\\2 c_{1} \nu_{2}^{2} e^{2 i \theta} & 0 & 2 V_{1} c_{2} e^{i \theta} & 0 & 2 c_{1} \mu_{2}^{2} & 0 & 0 & 0 & 0 & 0\\0 & 0 & 0 & 2 c_{1} \mu_{2}^{2} & 0 & 2 V_{1} c_{2} e^{i \theta} & 0 & 2 c_{1} \nu_{2}^{2} e^{2 i \theta} & 0 & 0\\0 & 0 & 2 c_{1} \nu_{2}^{2} e^{2 i \theta} & 0 & 2 V_{1} c_{2} e^{i \theta} & 0 & 2 c_{1} \mu_{2}^{2} & 0 & 0 & 0\\0 & 0 & 0 & 0 & 0 & 2 c_{1} \mu_{2}^{2} & 0 & 2 V_{1} c_{2} e^{i \theta} & 0 & 2 c_{1} \nu_{2}^{2} e^{2 i \theta}\\0 & 0 & 0 & 0 & 2 c_{1} \nu_{2}^{2} e^{2 i \theta} & 0 & 2 V_{1} c_{2} e^{i \theta} & 0 & 2 c_{1} \mu_{2}^{2} & 0\\0 & 0 & 0 & 0 & 0 & 0 & 0 & 2 c_{1} \mu_{2}^{2} & 0 & B'_{9,10} \\0 & 0 & 0 & 0 & 0 & 0 & 2 c_{1} \nu_{2}^{2} e^{2 i \theta} & 0 & B'_{10,9} & 0\end{matrix}\right],
\end{equation}
%
where $A'_{1,1} = A'_{10,10} = V_{1} \nu_{2}^{2} + \mu_{2}^{2}$, $A'_{2,2} = A'_{9,9} = V_{1} \mu_{2}^{2} + \nu_{2}^{2}$, and $B'_{1,2}=B'_{2,1}=B'_{9,10}=B'_{10,9}=c_{2} \left(V_{1} + 1\right) e^{i \theta}$.
Eqs.~(\ref{CM_A10}-\ref{CM_B10}) indicate that 
	each mode in time slot $n$ is correlated with 5 other modes in time slots $n-1$, $n$, and $n+1$ (with non-zero covariance), and is independent of modes outside these time slots (with zero covariance).
	When tracing out the boundary modes (1, 2, 9, and 10 for the special case of Fig.~\ref{fig_state}), we derive the covariance matrix without boundary modes, which corresponds to the quantum state discussed in our main text for $M=6$:
	\begin{equation}
		\label{CM_6} 
		\sigma^{(6)} = \left[\begin{matrix} \mathbf{A}^{(6)} & \mathbf{B}^{(6)} \\ (\mathbf{B}^{(6)})^* & (\mathbf{A}^{(6)})^* \end{matrix}\right],
	\end{equation}
	with
	\begin{equation}
		\label{CM_A6}
		\mathbf{A}^{(6)} = \left[\begin{matrix} V_{1} V_{2} & 0 & 2 c_{1} c_{2} e^{i \theta} & 0 & 0 & 0 \\ 0 & V_{1} V_{2} & 0 & 2 c_{1} c_{2} e^{- i \theta} & 0 & 0 \\2 c_{1} c_{2} e^{- i \theta} & 0 & V_{1} V_{2} & 0 & 2 c_{1} c_{2} e^{i \theta} & 0 \\ 0 & 2 c_{1} c_{2} e^{i \theta} & 0 & V_{1} V_{2} & 0 & 2 c_{1} c_{2} e^{- i \theta} \\0 & 0 & 2 c_{1} c_{2} e^{- i \theta} & 0 & V_{1} V_{2} & 0 \\0 & 0 & 0 & 2 c_{1} c_{2} e^{i \theta} & 0 & V_{1} V_{2}  \end{matrix}\right]
	\end{equation}
	and
	\begin{equation}
		\label{CM_B6}
		\mathbf{B}^{(6)} = \left[\begin{matrix}  
			0 & 2 V_{1} c_{2} e^{i \theta} & 0 & 2 c_{1} \nu_{2}^{2} e^{2 i \theta} & 0 & 0 \\
			2 V_{1} c_{2} e^{i \theta} & 0 & 2 c_{1} \mu_{2}^{2} & 0 & 0 & 0 \\
			0 & 2 c_{1} \mu_{2}^{2} & 0 & 2 V_{1} c_{2} e^{i \theta} & 0 & 2 c_{1} \nu_{2}^{2} e^{2 i \theta} \\
			2 c_{1} \nu_{2}^{2} e^{2 i \theta} & 0 & 2 V_{1} c_{2} e^{i \theta} & 0 & 2 c_{1} \mu_{2}^{2} & 0 \\
			0 & 0 & 0 & 2 c_{1} \mu_{2}^{2} & 0 & 2 V_{1} c_{2} e^{i \theta} \\
			0 & 0 & 2 c_{1} \nu_{2}^{2} e^{2 i \theta} & 0 & 2 V_{1} c_{2} e^{i \theta} & 0 \end{matrix}\right].
	\end{equation}
		By further calculations, in the case where the train of pump pulses of both OPAs exceeds the detected time slots, the covariance matrix for the quantum state out of an USUI with $M \in 2\mathbb{N}^+$ can be summarized as: 
	\begin{equation}
		\label{CM_s_def} 
		\sigma^{(M)} = \left[\begin{matrix} \mathbf{A}^{(M)} & \mathbf{B}^{(M)} \\ (\mathbf{B}^{(M)})^* & (\mathbf{A}^{(M)})^* \end{matrix}\right],
	\end{equation} 
	with non-zero elements of $\mathbf{A}^{(M)}$ and $\mathbf{B}^{(M)}$ given by
	\begin{equation}\label{As}
		\begin{aligned}
			A^{(M)}_{p,p} = & V_1V_2, \qquad p=1,2,...,M \\
			A^{(M)}_{p+2,p} = & 2 c_1 c_2 \mathrm{e}^{ i \theta (-1)^p },  \qquad p = 1,2,...,M-2 \\
			A^{(M)}_{p,p+2} = & 2 c_1 c_2 \mathrm{e}^{i \theta (-1)^{p+1} }, \qquad p = 1,2,...,M-2 
		\end{aligned}
	\end{equation}
	and
	\begin{equation}\label{Bs}
		\begin{aligned}
			B^{(M)}_{p+1,p} =  B^{(M)}_{p,p+1} = &
			2\sin^2(p\pi/2) V_1c_2\mathrm{e}^{i\theta} 
			+ 2\cos^2(p\pi/2) c_1\mu_2^2,  \qquad p = 1,2,...,M-1 \\ 
			B^{(M)}_{p+3,p} =  B^{(M)}_{p,p+3} = &
			2 \sin^2(p\pi/2) c_1 \nu_2^2 \mathrm{e}^{2i\theta}, \qquad p = 1,2,...,M-3 
		\end{aligned}
	\end{equation}
	respectively. The result in Eq.~(\ref{CM_s_def}-\ref{Bs}) are used to derive the $g^{(2)}$ and the $R_d^{(M)}$ in our main text.  
	
	\section{Derivation of the intensity correlation function $g^{(2)}$} \label{Sec2} 
	\setcounter{equation}{15}
	The photon statistics of the state out of USUI is decided by the average photon number $\langle \hat{N}_p \rangle = \langle \hat{A}^{\dagger}_p \hat{A}_p \rangle$ and the intensity covariance matrix $\mathbf{K}$ with elements defined as  $K_{p,q} \equiv  \langle \hat{N}_p \hat{N}_q \rangle - \langle \hat{N}_p \rangle \langle \hat{N}_q \rangle$, 
	which can be derived from $\bm{\alpha}$, $\mathbf{A}$ and $\mathbf{B}$ by Ref.\cite{Vallone2019PRA},
	\begin{align}
		\label{Nm}
		\langle \hat{N}_p \rangle = & \frac{1}{2}(A_{p,p}-1) + \langle \hat{A}_p \rangle\langle \hat{A}^{\dagger}_p\rangle 
		\\
		\label{KM}
		\mathbf{K} = &\frac{1}{4} ( \mathbf{A}\circ\mathbf{A}^* +  \mathbf{B}\circ\mathbf{B}^* - \mathbb{I}_M)  + 
		\mathbf{Re}\Big[ (\bm{\alpha^*} \bm{\alpha}^T) \circ \mathbf{A} + (\bm{\alpha^*} \bm{\alpha}^\dagger) \circ \mathbf{B}\Big].
	\end{align}
	The symbol $\circ$ denotes the Hadamard product of matrices, and $\mathbb{I}_M$ represents the $M \times M$ identity matrix.

	We consider vacuum input to OPA1, for which the displacement vector $\textbf{d}$ vanishes, i.e., $\langle \hat{A}_p \rangle=\langle \hat{A}^{\dagger}_p\rangle = 0$ for all $p$. 
	By substituting Eq. (\ref{CM_A6}-\ref{CM_B6}) (the $M=6$ case of Eq.~(\ref{As}-\ref{Bs})) into Eq.~(\ref{Nm}) and Eq.~(\ref{KM}), we obtain
	\begin{equation}
		\label{Nm_0}
		\langle \hat{N}_p \rangle =  \frac{1}{2}(V_1V_2-1),
	\end{equation}
	and
	\begin{equation}
		\mathbf{K}^{(6)} = \left[\begin{matrix}\frac{1}{4} V_{1}^{2} V_{2}^{2} - \frac{1}{4} & V_{1}^{2} c_{2}^{2} & c_{1}^{2} c_{2}^{2} & c_{1}^{2} \nu_{2}^{4} & 0 & 0\\V_{1}^{2} c_{2}^{2} & \frac{1}{4} V_{1}^{2} V_{2}^{2} - \frac{1}{4} & c_{1}^{2} \mu_{2}^{4} & c_{1}^{2} c_{2}^{2} & 0 & 0\\c_{1}^{2} c_{2}^{2} & c_{1}^{2} \mu_{2}^{4} & \frac{1}{4} V_{1}^{2} V_{2}^{2} - \frac{1}{4} & V_{1}^{2} c_{2}^{2} & c_{1}^{2} c_{2}^{2} & c_{1}^{2} \nu_{2}^{4}\\c_{1}^{2} \nu_{2}^{4} & c_{1}^{2} c_{2}^{2} & V_{1}^{2} c_{2}^{2} & \frac{1}{4} V_{1}^{2} V_{2}^{2} - \frac{1}{4} & c_{1}^{2} \mu_{2}^{4} & c_{1}^{2} c_{2}^{2}\\0 & 0 & c_{1}^{2} c_{2}^{2} & c_{1}^{2} \mu_{2}^{4} & \frac{1}{4} V_{1}^{2} V_{2}^{2} - \frac{1}{4} & V_{1}^{2} c_{2}^{2}\\0 & 0 & c_{1}^{2} \nu_{2}^{4} & c_{1}^{2} c_{2}^{2} & V_{1}^{2} c_{2}^{2} & \frac{1}{4} V_{1}^{2} V_{2}^{2} - \frac{1}{4}\end{matrix}\right],
	\end{equation}
	which is enough to illustrate the pattern of $\mathbf{K}$ for mode numbers larger than 6: it can be viewed as three $2\times2$ matrix $\mathbf{K_{diag}} = \left[\begin{matrix} \frac{1}{4} V_{1}^{2} V_{2}^{2} - \frac{1}{4} & V_{1}^{2} c_{2}^{2}\\ V_{1}^{2} c_{2}^{2} & \frac{1}{4} V_{1}^{2} V_{2}^{2} - \frac{1}{4} \end{matrix}\right]$, 
	$\mathbf{K_1}=\left[\begin{matrix} c_1^2 c_2^2 & c_1^2 \nu_2^4 \\ c_{1}^{2} \mu_{2}^{4} & c_1^2 c_2^2 \end{matrix}\right]$, and 
	$\mathbf{K_2}=\mathbf{K_1^T}$ periodically filled into a large zero matrix. 
	The non-zero elements of $\mathbf{K}$ for an arbitrary mode number $M$ are given by
	\begin{equation}\label{Ks_M}
		\begin{aligned}
			K^{(M)}_{p, p}  =&
			\frac{1}{4}(V_1^2V_2^2-1), \qquad p = 1, 2, ..., M ~,\\
			K^{(M)}_{p+1, p} =& K^{(M)}_{p, p+1}  = 	\sin^2(p\pi/2) V_1^2c_2^2 + \cos^2(p\pi/2) c_1^2\mu_2^4,  \qquad  p = 1, 2, ..., M-1 ~,\\
			K^{(M)}_{p+2, p} =& K^{(M)}_{p, p+2}  = 	c_1^2c_2^2, \qquad p = 1, 2, ..., M-2 ~,\\
			K^{(M)}_{p+3, p} =& K^{(M)}_{p, p+3}  = 	\sin^2(p\pi/2) c_1^2\nu_2^4,   \qquad  p = 1, 2, ..., M-3 ~. 
		\end{aligned}
	\end{equation}
	%
	
	The intensity correlation function $g^{(2)}(p,q)$ between modes $p$ and $q$ is related to the photon number covariance matrix $\mathbf{K}$ through
	\begin{equation}\label{eq:g_pq}
		g^{(2)}(p,q) = 
		\frac{\langle\hat{A}_p^{\dagger}\hat{A}_q^{\dagger}\hat{A}_p\hat{A}_q\rangle}{\langle \hat{A}_p^{\dagger} \hat{A}_p \rangle \langle \hat{A}_q^{\dagger} \hat{A}_q\rangle}
		=\frac{K_{p,q}}{\langle \hat{N}_p \rangle \langle \hat{N}_q \rangle}-\frac{\delta_{p,q}}{\langle \hat{N}_p \rangle}+1.
	\end{equation}
	Using the calculated results in Eq.~(\ref{Nm_0}) and Eq.~(\ref{Ks_M}), we obtain $g^{(2)}$ for an arbitrary mode pair, as summarized in Table 1 of the main text.
	
	\section{Derivation of multi-mode intensity squeezing $R_d^{(M)}$}\label{Sec3}
	\setcounter{equation}{21}
	We consider a coherent-state input at the signal port of OPA1, with the idler port remaining in vacuum. In this case, the initial displacement vector of the signal mode has nonzero elements $ x /\sqrt{2} $ with $x \gg 1$. Consequently, the displacement vector $\textbf{d}$ is modified, while the covariance matrix $\sigma^{(M)}$ (Eq.~(\ref{CM_s_def}-\ref{Bs})) remains unchanged.
	By following the same computational procedure as in Appendix~\ref{Sec1} and Appendix~\ref{Sec2}, we can obtain 
	\begin{align}\label{a_Nm}
		\langle \hat{N}_p \rangle =   \frac{1}{4}x^2\left[2(\mu_1\mu_2+\nu_1\nu_2)^2-1+(-1)^{(p+1)}\right], 
	\end{align}
	and $\mathbf{K}$ with $p = 1, 2, ..., M$. 
	In the calculation, we retain only the $x^2$ term and neglect the lower-order terms, since $x \gg 1$.
	The phase difference of the pumps is set to $\theta=0$.

	Using the definition of covariance matrix $\mathbf{K}$, the variance of  $\sum_{p=1}^M w_p \hat{N}_p$ can be expressed as a quadratic form in the coefficients of the linear combination:
	\begin{eqnarray}
		\label{var_lin_N}
		\Delta( \sum_{p=1}^M w_p \hat{N}_p )^2  = \mathbf{w} \mathbf{K} \mathbf{w}^T,
	\end{eqnarray}
	where $\mathbf{w} = [w_1, w_2, \cdots, w_M]$ is the parameter vector for the linear combination, $M$ is even, and the square variance for a given operator $\hat O$ is defined as $\Delta\hat O ^2 = \langle \hat O ^2 \rangle - \langle\hat O\rangle ^2$. 
	Eq.~(\ref{var_lin_N}) can be normalized to the shot-noise limit (SNL) to evaluate the multi-mode intensity squeezing level: 
	\begin{equation}\label{Rt_e}
		R_d^{(M)} = \frac{\Delta( \sum_{p=1}^M w_p \hat{N}_p )^2}{\sum^M_{p=1} |w_p| \langle \hat{N}_p\rangle} = \frac{\mathbf{w} \mathbf{K} \mathbf{w}^T}{\sum^M_{p=1} |w_p| \langle \hat{N}_p\rangle}.
	\end{equation}
	Let's first derive the variance of the intensity-difference between two modes of the quantum system $\{-1s,-1i,0s,$ $0i,1s,1i\}$ shown in Fig.~1(b) of the main text.
	Using Eq.~(\ref{var_lin_N}), $\Delta( \sum_{p=1}^{6} w_p \hat{N}_p )^2  = \mathbf{w} \mathbf{K} \mathbf{w}^T$, the two-mode intensity-difference variances for the mode pairs $\{0s,0i\}$, $\{0s,1s\}$, $\{0s,-1s\}$, $\{0s,1i\}$, $\{0s,-1i\}$, $\{-1s,1s\}$ and $\{-1s,1i\}$ are obtained by choosing 
	$\mathbf{w}=[0,0,1,-1,0,0]$, 
	$[0,0,1,0, \allowbreak -1,0]$, 
	$[-1,0,1,0,0,0]$,
	$[0,0,1,0,0,-1]$, $[0,-1,1,0,0,0]$,
	$[1,0,0,0,-1,0]$, and 
	$[1,0,0,0,0,-1]$, respectively.
	These representative cases allow the results for all possible mode pairs of the multi-mode quantum state out of the USUI to be obtained straightforwardly.
	For comparison, we first present the normalized intensity noise of an individual mode:
	\begin{equation}\label{Rt_single}
		R_d^{(1)} = (\mu_1^2+\nu_1^2)(\mu_2^2+\nu_2^2).
	\end{equation}
	Then, we take $0s$ as the reference mode and summarize the normalized two-mode intensity-difference noise between mode $0s$ and an arbitrary mode as follows:
	\begin{eqnarray}
		\label{Rd34}
		R_d^{(2)}(0s,0i) 
		&= & \frac{4(\nu_1^4+\nu_1^2)+1}{2(\mu_1\mu_2+\nu_1\nu_2)^2-1} ,\\
		\label{Rd35}
		R_d^{(2)}(0s,1s) 
		&= & (\mu_1^2+\nu_1^2)(\mu_2^2+\nu_2^2)-2\mu_1\mu_2\nu_1\nu_2,
		\\
		\label{Rd13}
		R_d^{(2)}(0s,-1s) 
		&= & (\mu_1^2+\nu_1^2)(\mu_2^2+\nu_2^2)-2\mu_1\mu_2\nu_1\nu_2,
		\\
		\label{Rd36}
		R_d^{(2)}(0s,1i) 
		&= & (\mu_1^2+\nu_1^2)(\mu_2^2+\nu_2^2) 
		- 2\mu_1\nu_1\nu_2^2 \frac{2\mu_1\nu_1(\mu_2^2+\nu_2^2)+2\mu_2\nu_2(\mu_1^2+\nu_1^2)}{2(\mu_1\mu_2+\nu_1\nu_2)^2-1} ,\\
		\label{Rd45}
		R_d^{(2)}(0s,-1i) 
		&= & (\mu_1^2+\nu_1^2)(\mu_2^2+\nu_2^2) 
		- 2\mu_1\nu_1\mu_2^2 \frac{2\mu_1\nu_1(\mu_2^2+\nu_2^2)+2\mu_2\nu_2(\mu_1^2+\nu_1^2)}{2(\mu_1\mu_2+\nu_1\nu_2)^2-1} ,\\
		%
		%
		\label{Rd38}
		R_d^{(2)}(0s,q) 
		&= & (\mu_1^2+\nu_1^2)(\mu_2^2+\nu_2^2), ~~~ q \notin \{0s, 0i, \pm1s,\pm1i \}.
	\end{eqnarray}
	For the special case $\mu_1=\mu_2=\mu\gg1$, i.e., equal-gain OPAs operating in the high-gain regime, 
	Eqs.~(\ref{Rt_single}–\ref{Rd38}) reduce to
	\begin{eqnarray}
		\label{Rd-1}    
		R_d^{(1)}        =  (\mu^2+\nu^2)^2 ~~~~~  &\stackrel{\mu\gg1}\Longrightarrow& ~~~~~ 4\mu^4,\\
		\label{Rd34-}
		R_d^{(2)}(0s,0i) =  \frac{4(\nu^4+\nu^2)+1}{2(\mu^2+\nu^2)^2-1} ~~~~~  &\stackrel{\mu\gg1}\Longrightarrow& ~~~~~ \frac{1}{2},\\
		\label{Rd35-}
		R_d^{(2)}(0s,1s)  =  (\mu^2+\nu^2)^2-2\mu^2\nu^2 ~~~~~  &\stackrel{\mu\gg1}\Longrightarrow& ~~~~~ 2\mu^4,\\
		\label{Rd13-}
		R_d^{(2)}(0s,-1s)  =  (\mu^2+\nu^2)^2-2\mu^2\nu^2 ~~~~~  &\stackrel{\mu\gg1}\Longrightarrow& ~~~~~ 2\mu^4,\\
		\label{Rd36-}
		R_d^{(2)}(0s,1i)  =  (\mu^2+\nu^2)^2 - 4\mu^2\nu^4 \frac{(\mu^2+\nu^2)}{(\mu^2+\nu^2)^2-1}  ~~~~~  &\stackrel{\mu\gg1}\Longrightarrow& ~~~~~ 2\mu^4,\\
		\label{Rd45-}
		R_d^{(2)}(0s,-1i)  =  (\mu^2+\nu^2)^2 - 4\mu^4\nu^2 \frac{(\mu^2+\nu^2)}{(\mu^2+\nu^2)^2-1}  ~~~~~  &\stackrel{\mu\gg1}\Longrightarrow& ~~~~~ 2\mu^4,\\
		%
		%
		\label{Rd38-}
		R_d^{(2)}(0s,q)  =  (\mu^2+\nu^2)^2, ~~~ q \notin \{0s, 0i, \pm1s,\pm1i \}  ~~~~~  &\stackrel{\mu\gg1}\Longrightarrow& ~~~~~ 4\mu^4.
	\end{eqnarray}
	%
	From Eqs.~(\ref{Rd-1}–\ref{Rd38-}), 
	we find that $R_d^{(2)}(0s,0i)$, $R_d^{(2)}(0s,\pm 1s)$ and $R_d^{(2)}(0s,\pm 1i)$ are lower than $R_d^{(1)}$, indicating that mode $0s$ is correlated with modes $0i$, $1s$, $-1s$, $1i$, and $-1i$.
	The correlations between mode $0s$ and modes $1s$, $-1s$, $1i$, and $-1i$ reduce the normalized intensity-difference noise from $4\mu^4$ (uncorrelated case) to $2\mu^4$. For mode pairs within the same time slot, e.g., $(0s,0i)$, stronger correlations further suppress the normalized noise to $1/2$, below the shot-noise level.
	To further enhance the level of intensity squeezing, it is necessary to measure more signal and idler modes within the same time slots. 
	We therefore perform a linear combination of $M/2$ signal modes and $M/2$ idler modes, i.e., $\sum_{p=1}^M w_p \hat{N}_p$, where each signal–idler pair occupies the same time slot. 
	The variance of $\sum_{p=1}^M w_p \hat{N}_p$ is then given by
	\begin{equation}\label{wkw}
		\Delta( \sum_{p=1}^M w_p \hat{N}_p )^2 = \mathbf{w} \mathbf{K} \mathbf{w}^T = 2x^2\nu_1^4 + 2x^2\nu_1^2 + \frac{M}{4}x^2, 
	\end{equation}
	with $\mathbf{w} = [\underbrace{1, -1, 1, -1, \cdots}_M]$. 
	Substituting Eq.~(\ref{a_Nm}) and Eq.~(\ref{wkw}) into Eq.~(\ref{Rt_e}) then yields the corresponding result:
	\begin{equation}\label{Rt}
		R_d^{(M)} = \frac{8(\nu_1^4 + \nu_1^2)}{M\left[2(\mu_1\mu_2+\nu_1\nu_2)^2-1\right]} + \frac{1}{2(\mu_1\mu_2+\nu_1\nu_2)^2-1}.
	\end{equation}
	Note that when the two OPAs operate with identical gains in the high-gain regime ($\mu_1^2=\mu_2^2=\mu^2\gg 1$), Eq.~(\ref{Rt}) reduces to the expression given in Eq.~(5) of the main text.
	It's straightforward to see that the first term in Eq.~(\ref{Rt}) vanishes when the number of measured modes $M$ becomes sufficiently large. In this limit, the intensity-difference squeezing reduces to 
	\begin{equation}
		R_d^{(M\gg1)} = \frac{1}{2g-1}, 
	\end{equation}
	where $g=(\mu_1\mu_2+\nu_1\nu_2)^2$ denotes the total intensity gain of the USUI.
	For a single OPA with gain $g$, the intensity difference squeezing degree is given by $R_d=1/(2g-1)$~\cite{Yan2025OL}. A balanced SU(1,1) interferometer (SUI) can be regarded as an effective OPA with an overall gain of $g=(\mu_1\mu_2+\nu_1\nu_2)^2$.
	Consequently, when $M\gg1$, the intensity difference squeezing of an USUI approaches that of the state out of a balanced SUI.

	\section{Detailed experimental setup}\label{Sec4}
	\begin{figure*}[t]
		\centering
		\includegraphics[width=0.9\textwidth]{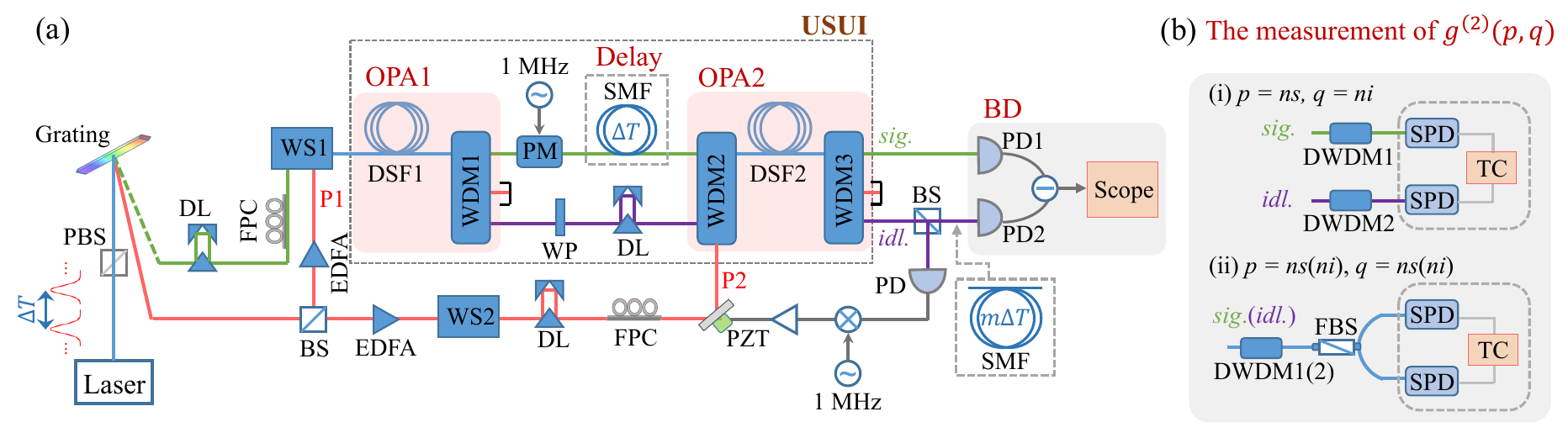}
		\caption{\label{Fig-setup-supp}Experimental setup. (a) An unbalanced SU(1,1) interferometer (USUI) based on an ultrashort-pulse-pumped fiber optical parametric amplifier (OPA) with fast balanced photocurrent detection(BD) for measuring multi-mode intensity squeezing. (b) Coincidence-counting setup used to measure intensity correlation function $g^{(2)}(p,q)$. Laser, mode-locked fiber laser; PBS, polarization beam splitter; DL, delay line; FPC, fiber polarization controller; WS, wave shaper; BS, beam splitter; EDFA, erbium-doped fiber amplifier; DSF, dispersion-shifted fiber; WDM, wavelength-division multiplexer; PM, phase modulator; SMF, single mode fiber; WP, Q–H–Q wave-plate sequence; PZT, piezoelectric transducer; PD, photo diode; DWDM, dense wavelength-division multiplexer; FBS, fiber beam splitter; SPD, single photon detector; TC, time controller.}
	\end{figure*}
	The detailed experimental setup is shown in Fig.~\ref{Fig-setup-supp}. 
	The SUI is formed by two fiber optical parametric amplifiers (OPAs), which are based on four wave mixing (FWM) in dispersion-shifted fibers (DSFs)~\cite{agrawal2000nonlinear}. 
	An USUI is realized by inserting a section of standard single-mode fiber (SMF) into the inter-stage signal path between the two OPAs.
	OPA1 consists of a DSF and a wavelength division multiplexer (WDM); while OPA2 consists of a DSF and two WDMs. 
	The two DSFs are identical 
	with lengths of $\sim 300$ m.
	Their zero-dispersion wavelength, nonlinear coefficient and dispersion slope are $\lambda_0\approx 1548.5$ nm, $\gamma \approx 2~(\mathrm{W \cdot km})^{-1}$, and $D_{slope} \approx 0.075~\mathrm{ps/(km \cdot nm^2)}$, respectively. 
	When the phase matching condition $\Delta k \approx (-k^{(2)}/4)(\omega_s-\omega_i)^2-2\gamma P_p\approx 0$ is satisfied in each DSF~\cite{agrawal2000nonlinear}, efficient FWM occurs in both DSF1 and DSF2, resulting in a gain bandwidth of up to 40 nm~\cite{li2007spectral}. 
	In this process, two pump photons at frequency $\omega_p$ are annihilated simultaneously, generating a pair of signal and idler photons at $\omega_s$ and $\omega_i$, respectively. The process satisfies energy and momentum conservation, and the generated photons exhibit correlations in frequency, time, energy, etc.  
	Here, $k^{(2)}=-(\lambda_p^2/2\pi c)D_{slope}(\lambda_p-\lambda_0)$ denotes the second order dispersion coefficient of the DSFs. 
	$\lambda_p$ and $P_p$ are the central wavelength and peak power of the pump, respectively. $c$ is the speed of light in vacuum. When $\lambda_p > \lambda_0$, i.e., in the anomalous dispersion regime, $k^{(2)} < 0$, leading to $\Delta k \approx 0$.
	To suppress Raman noise in the fibers~\cite{li2004all, lee2006generation}, the DSFs are immersed in liquid nitrogen (77 K). 
	WDM1 and WDM3 consist of two coarse wavelength-division multiplexer filters, centered at 1530 nm and 1570 nm, respectively, each with a $\pm 7.5$ nm passband at the -0.3 dB level.
	In contrast, WDM2 includes an additional channel centered at 1550 nm with the same passband.
	The WDMs are employed to separate or combine the pump, signal, and idler fields.

\begin{figure}[h]
	\centering
	\includegraphics[width= 0.35\textwidth]{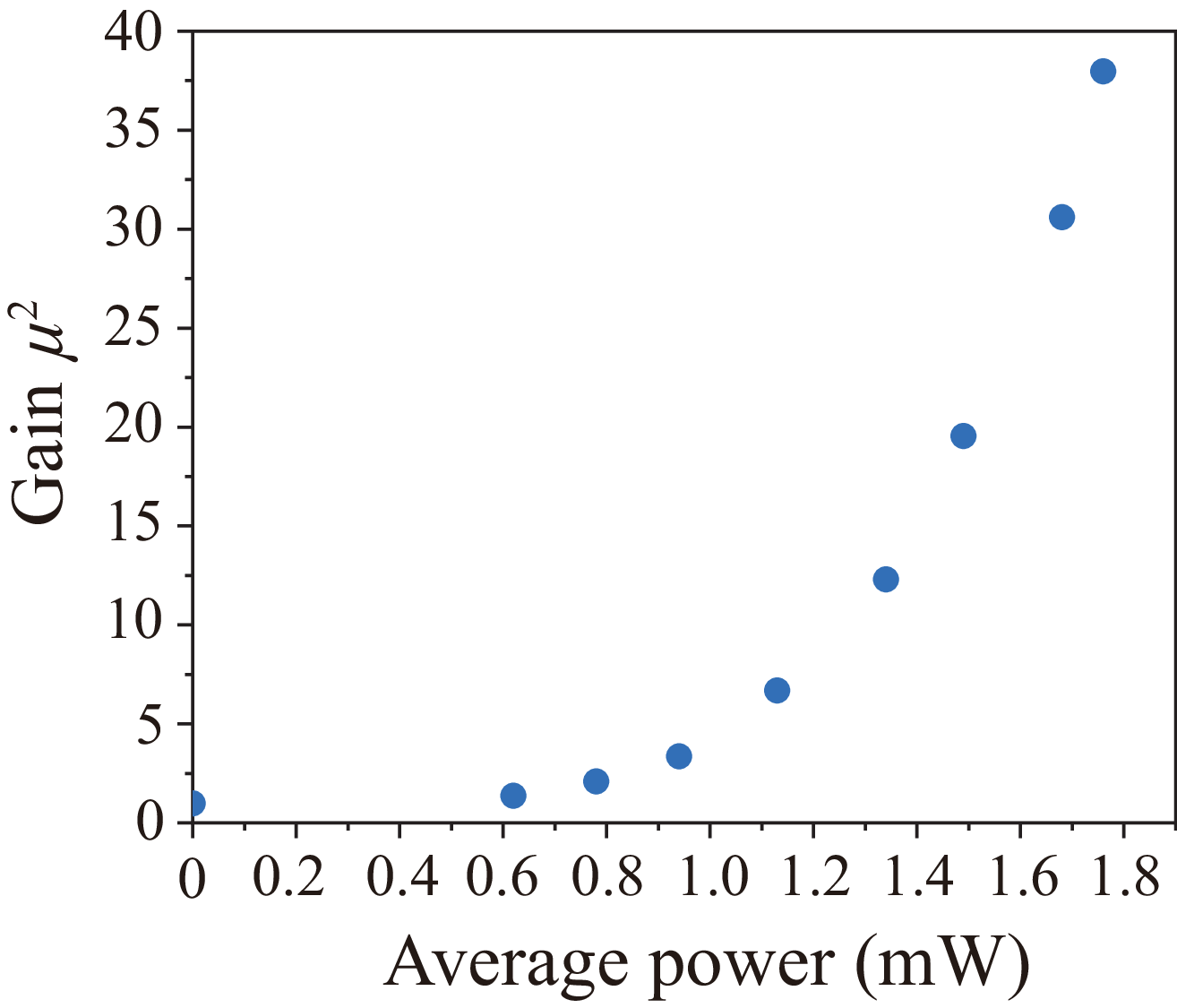}
	\caption{\label{Fig-gain} Power gain $\mu^2$ of the OPA versus average pump power for an injected signal power of $I_s^{in}\approx 1~\mu\mathrm{W}$.}
	\end{figure}

	The pulsed pump (red) and seed (green) fields are obtained by filtering the output of a mode-locked fiber laser, which has a repetition rate of 50 MHz, a central wavelength of 1550 nm, and a 5~dB bandwidth of 40 nm.
	Depending on the experimental configuration, the seed field can either be injected or blocked.
	The pulsed pump is split by a beam splitter (BS) into two paths, P1 and P2, which serve as the pump fields for OPA1 and OPA2, respectively.
	P1 is amplified by an erbium-doped fiber amplifier (EDFA) and then combined with the seed (when it is enabled) using a wave shaper (WS, Finisar 4000A).
	The polarization and temporal modes of the pump and seed are matched using a fiber polarization controller (FPC) and a delay line (DL).
	The transmissivity and spectral profiles of the pump and seed fields are adjusted by the WS, where both spectra are shaped into Gaussian profiles with FWHM of 0.45 nm for the pump and 1.3 nm for the seed.
	Meanwhile, an additional EDFA and WS amplify P2 and shape its spectrum. 
	The central wavelengths of P1 and P2 are set to 1549.3 nm to satisfy the phase-matching condition for FWM in the DSFs. 
	The injected seed is set to a center wavelength of 1533 nm.
	When the seed is blocked, spontaneous FWM occurs in DSF1, 
	corresponding to a vacuum seed $|0\rangle$ injected into OPA1 in Fig.~1(a) of the main text.
	In contrast, when the seed is injected, the process becomes stimulated FWM. 
	This corresponding to a coherent seed $|\alpha \rangle$ injected into OPA1 in Fig.~1(a) of the main text.
	In this process, the injected seed, serving as the signal field, is amplified by the pump, generating a pulsed idler field centered at 1566 nm.
	The power gain of the OPA, $\mu^2=I_s^{out}/I_s^{in}$, defined as the ratio of the amplified to injected signal power, is plotted versus the average pump power in Fig.~\ref{Fig-gain}.
	
	The signal and idler fields generated by the FWM process in DSF1 are first spatially separated using WDM1.
	To satisfy the FWM condition in DSF2, two delay lines (DLs), a Q–H–Q wave-plate sequence, and a FPC are employed to match the temporal modes and polarization of the pump, signal and idler fields at the input of DSF2.
	In addition, a phase modulator (PM) is placed in the signal arm to apply a controlled modulation for phase locking, while a $\sim4.08$-m-long section of SMF is introduced to provide a temporal delay of one pulse period. 
	The signal and idler fields are then combined with P2 using WDM2 and launched into DSF2.
	At the output of DSF2, they are separated by WDM3 and routed to the detection system.
	
	%
	We use self-homodyne detection to measure multimode intensity squeezing, for which the USUI needs to be phase locked at the point of maximum output intensity, corresponding to a pump phase difference of $\theta = 0$.
	This is achieved by applying a 1-MHz phase modulation to the inter-stage signal field via a PM.
	A small fraction (1\%) of the idler field at the output of the USUI is tapped using a 99:1 BS and detected by a photodetector. The detected signal is then demodulated by frequency mixing to generate an error signal, which is processed by a Proportional-Integral-Differential (PID) controller and fed back to the piezoelectric transducer (PZT) in the path of P2. This feedback loop stabilizes the relative phase of the interferometer, thereby achieving phase locking of the USUI.
	The intensity difference noise of the output signal and idler (taken from the 99\% output port of the 99:1 BS) fields are measured by using the fast balanced photodetection (BD) system consisting of two photo diodes (PDs) and related electronic circuits (not shown in Fig.~\ref{Fig-setup-supp}). The photocurrent difference of the two PDs is recorded by a scope with a voltage resolution of 12 bits.

	\begin{figure*}[h]
		\centering
		\includegraphics[width=0.7\textwidth]{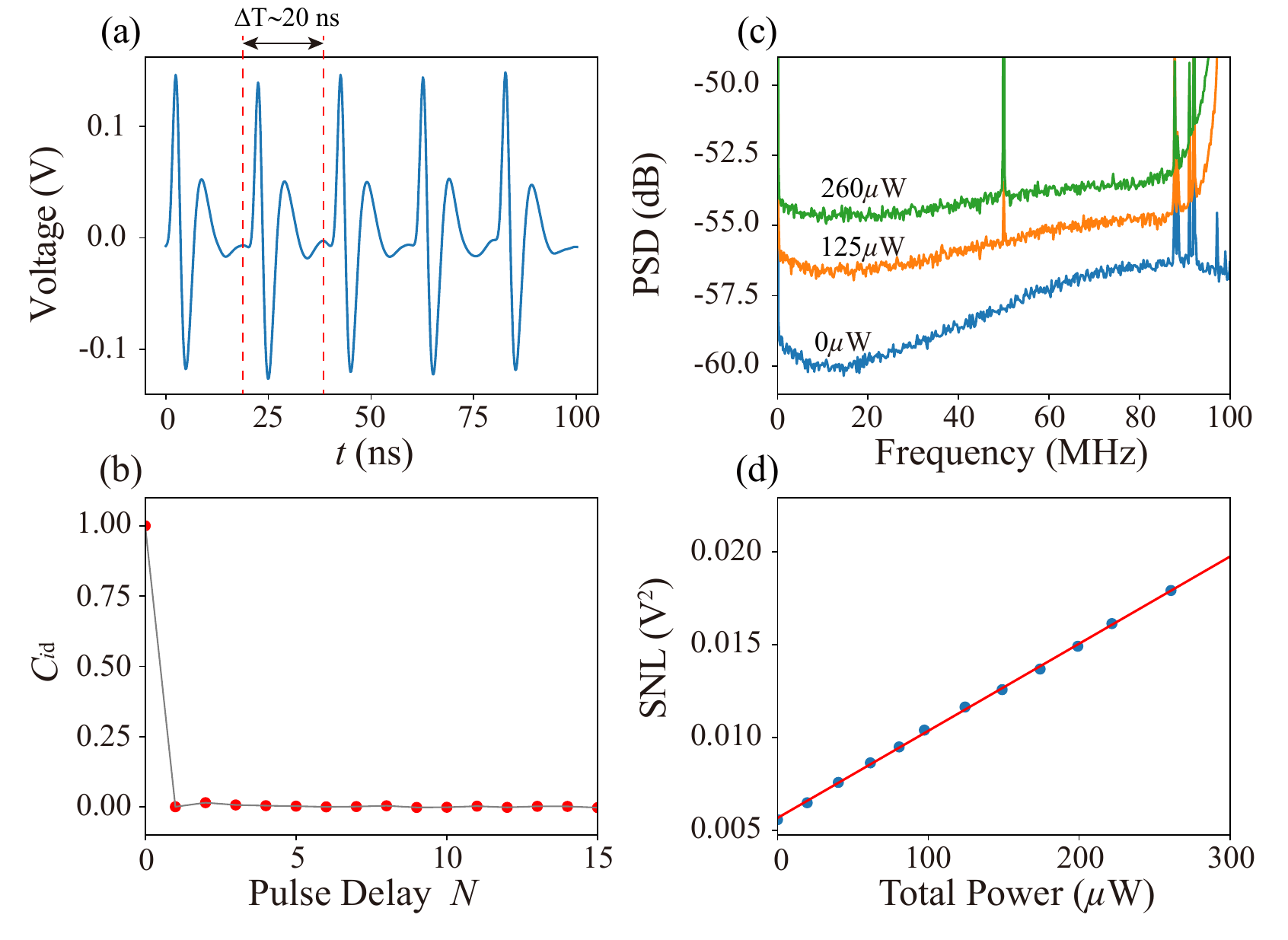}
		\caption{\label{Fig-HD-cal}Calibration results for the fast balanced photocurrent detection (BD) system\cite{Zhao2023OL}. (a) Voltage trace recorded by the digital oscilloscope over five pulse periods. (b) Normalized correlation coefficient $C_{id}$ between the $n$th and $(n+N)$th pulse-integrated voltages measured with the fast BD. (c) Shot-noise power spectral density measured under different total optical input powers. (d) Shot-noise level of the BD calibrated at various optical input powers, where the red line represents a linear fit to the data points.}
	\end{figure*}
	
	A schematic of the intensity correlation function $g^{(2)}(p,q)$ detection setup is separately shown in Fig.~\ref{Fig-setup-supp}(b).
	The signal and idler fields at the output of the USUI are filtered by two 100-GHz dense wavelength-division multiplexers (DWDMs), with DWDM1 and DWDM2, respectively, providing more than 120 dB of isolation from the pump light.
	We note that two cases are considered in the measurement of $g^{(2)}(p,q)$: (i) the modes $p=ns$ and $q=ni$ in different (signal and idler, respectively) channels, and (ii) the modes $p=ns(ni)$ and $q=ns(ni)$ in same (signal or idler) channel.
	For case (i), the output of DWDM1 and DWDM2 are sent to two superconducting nano-wire single photon detectors (SPDs). 
	For case (ii), a 50:50 fiber beam splitter (FBS) is placed after DWDM1 or DWDM2 to enable a Hanbury Brown and Twiss-type measurement.
	For both cases (i) and (ii), the electrical signals produced by the SPDs in response to incoming photons are reshaped and recorded by an Time Controller (TC, ID Quantique SA ID1000 Time Controller).
	As the TC records all detection events, the single-count rates of the two SPDs, $R_1$ and $R_2$, as well as the two-fold coincidence rate $R_c$, can be extracted from different time bins. The measured intensity correlation function $g^{(2)}$ is given by $g^{(2)}=R_c/R_{acc}=R_c/(R_1R_2/R_p)$, where $R_p=50$ MHz denotes the pulse repetition rate of the detected fields.
	To reduce measurement uncertainty, each set of single-channel and coincidence counts was accumulated over an integration time of 200 s to evaluate $g^{(2)}$.

	\section{Calibration of the fast balanced photocurrent detection system}\label{Sec5}
	P1 (coherent state) is split by a 50:50 BS and used to calibrate the fast BD, following the method reported in Ref.~\cite{Zhao2023OL}. 
	Fig.~\ref{Fig-HD-cal}(a) presents a representative output waveform of the BD measured by the scope, in which each detected pulse pair produces a corresponding voltage pulse.
	We identify each voltage pulse (indicated by the red dashed line in Fig.~\ref{Fig-HD-cal}(a)) and sum the acquired data points over a duration of $\Delta T=20$ ns. The resulting value, denoted as $e_n$, is used as an estimate of the intensity difference between the optical pulse pair simultaneously sent to the BD in the $n$th time slot.
	For multi-mode intensity squeezing measurements, the sum of $e_n$ over multiple time slots can be used to characterize the intensity difference between multiple optical pulse pairs.
	The calculated normalized correlation coefficient $C_{id}$ between the calculated $e_n$ values and its $N$-pulses-shifted values $e_{n-N}$ are shown by the red circles in Fig.~\ref{Fig-HD-cal}(b).
	The shot noise power spectrum density (PSD) of the balanced detector output, averaged over 250,000 time slots, are shown in Fig.~\ref{Fig-HD-cal}(c).
	The PSDs indicate that the BD has a bandwidth exceeding 80 MHz, which is lager than the repetition rate of the laser.
	The results shown in Fig.~\ref{Fig-HD-cal}(a)(b)(c) indicate that our fast BD can measure the optical pulses almost independently.
	We also calibrated the shot-noise level (SNL) of the BD at different optical powers, as shown in Fig.~\ref{Fig-HD-cal}(d), which exhibits good linear relationship between the optical power and the variance of the integrated voltage pulse.

\bibliography{main}

\end{document}